\documentclass{article}

\usepackage{amsmath,amssymb,epsf,a4wide}

\begin{document}
\newcommand{\nd}[1]{/\hspace{-0.5em} #1}

\begin{titlepage}

\begin{centering}
%

\vspace{0.1in}

{\Large {\bf Putting Liouville String Models of (Quantum) Gravity to Test}}

\vspace{0.4in}

{\bf Elias Gravanis and Nick E. Mavromatos } \\
\vspace{0.2in}
Department of Physics, Theoretical Physics, King's College London,\\
Strand, London WC2R 2LS, United Kingdom.

\vspace{0.4in}
 {\bf Abstract}

\end{centering}

\vspace{0.2in}

{\small Critical String Theory is by definition an $S$-matrix
theory. In this sense, (quantum) gravity situations where
a unitary $S$-matrix may not be a well-defined concept, as a consequence
of the existence of macroscopic (global) or 
microscopic (local) gravitational 
fluctuations with event horizons, present a challenge to string theory. 
In this article, we take some modest steps in suggesting alternative 
treatments of such cases via non-critical (Liouville) strings,
which do not have a well-defined $S$ matrix, but for which 
a superscattering \$ matrix is mathematically consistent.    
After a brief review of the underlying mathematical 
formalism, we consider a specific stringy model 
of induced non-criticality, with dynamical formation of
horizons, 
associated with the 
recoil of a $D$-particle defect, embedded in our 
four-dimensional space time, 
during its scattering with a (macroscopic) number of closed string states. 
We study in 
detail the associated spacetime 
distortion in the neighbourhood of the defect, 
which has the 
form of a finite-radius curved `bubble', 
matched with a 
Minkowskian space-time in the exterior. 
As a consequence of the non-criticality of the 
underlying $\sigma$-model, the space time 
is unstable, 
and has non-trivial stochastic properties: 
thermal properties due to 
its ``Rindler accelerating nature'', and entropy growth for an asymptotic 
observer, associated with 
information being carried away by the `recoil' degrees of freedom.   
For the validity of our approach 
it is essential that the string length is a few orders of magnitude 
larger than the Planck length, 
which is  a typical situation encountered
in many $D$-brane-world models.
An interesting feature of our model
is the emission of high-energy 
photons from the unstable bubble. 
As a result, the neighbourhood of the recoiling $D$-particle 
may operate as a source region of ultra-high-energy 
particles, which could reach the observation 
point if the source lies within the respective mean-free paths.
This may 
lead to severe phenomenological constraints on the model, e.g. 
in connection with 
apparent ``violations'' of the GZK cutoff, which we 
discuss briefly. We also comment on the effects  
of embedded recoiling $D$-particles on the cosmological evolution of 
a Friedman-Robertson-Walker type 
Universe. Specifically, we argue in favour of a relaxing-to-zero
vacuum energy a l\'a quintessence, and an eventual 
removal of global cosmological horizons, and hence stopping of the 
cosmic acceleration in the far future in such models.}

\vspace{0.8in}
\begin{flushleft}
July 2001 
\end{flushleft}

\end{titlepage}

\section{Introduction and Summary}

The field of critical string theory 
is a vast, and well established, field of modern 
theoretical physics, where an enormous intellectual 
effort has been devoted for the past twenty years.
Since the anomaly cancellation mechanism of Green and Schwarz~\cite{GSW},
which opened up the way for realistic phenomenology of string theory, 
a considerable amount of very attractive work 
has been provided, dealing with 
various aspects of string theories, ranging from 
low-energy physics to non-perturbative aspects. 
The recent discovery of string dualities and 
the associated solitonic structures of `string theory',
the so-called (D)irichlet branes~\cite{branes},
has enlarged the spectrum of applicability 
of string theory to important issues 
of quantum gravity that      
could 
not have been tackled before, 
such as the counting of microstates of certain species of 
black holes.  

However, despite this considerable effort, there are still 
important conceptual issues of quantum gravity which
are not answered by critical string theory, at least in our opinion.  
One of them is the compatibility of quantum coherence with 
the existence of singular space time configurations, such as 
black holes, which exist as classical solutions of General Relativity,
and are therefore expected to appear (at a microscopic Planckian
size) also in the appropriate (and yet-to-be discovered) measure 
of  gravitational path integration. 
Finding a proper treatment of spacetime singularities, 
encountered in general relativity, within the context of a unitary 
quantum theory still remains, in our opinion, 
a challenge for critical string theory.  Such a feeling is also supported
by the fact that, at present, 
a consistent quantization of eternally accelerating 
Universes is also poorly understood in this context~\cite{accel,emnsmatrix}.

These problems are associated with the fact that in 
such situations 
there are space-time boundaries (microscopic or macroscopic)
which prevent a proper definition of a scattering $S$ matrix.
Since  critical string theory is by construction 
a theory of an $S$ matrix, this presents an incompatibility. 
One way out of this impasse, therefore, would be to somehow enlarge the 
definition of `string theory' so that critical string theory is only a 
limiting case, and the full theory admits consistent quantization
but not necessarily through an $S$ matrix. 

A possible resolution of this important issue 
may be provided by an appropriate extension of a concept 
already encountered in local field theory of open systems.
The idea utilizes the concept of the so-called
superscattering or \$ matrix, which was suggested by 
Hawking~\cite{hawking} as a way of making sense of theories of 
quantum gravity. 
Taking into account that any complete theory of quantum gravity should
incorporate microscopic space-time singularities
(e.g. black holes) 
in the appropriate path-integral measure over the gravitational field, 
Hawking argued that 
such configurations appear to be 
incompatible, in general, with the maintenance 
of quantum coherence of matter theories interacting with them. 
Indeed, once such 
configurations are formed from the quantum-gravity `vacuum', 
they can `capture' part of an {\it initially pure} matter quantum state,
inside their microscopic event horizons.
Since a black hole {\it evaporates} in quantum theory, 
according to the celebrated
Hawking radiation effect, 
the relevant information appears be `lost'. Thus, an observer 
in the asymptotic future, and far away from 
the black hole, will necessarily observe a {\it mixed} 
quantum state, in analogy with the case of {\it open} quantum mechanical 
systems~\cite{ehns}. This evolution, from an initially pure quantum 
matter state to a mixed asymptotic
state in the presence of microscopic quantum-gravity vacuum 
singular configurations, results inevitably in the impossibility of 
treating the quantum gravity vacuum in a standard unitary 
field theoretic 
way. 
The \$ matrix relates asymptotic density matrix states, which are 
not pure states in general. In open systems, where there is 
`loss of information', an asymptotic observer should trace over 
such states, $\{ {\cal M} \}$, 
to obtain a density matrix $\rho ={\rm Tr}_{\cal M} |\Psi><\Psi|$. 
The \$ matrix is then defined:
\begin{equation}\label{supscat}  
\rho_{\rm out} = \$ \rho_{\rm in}
\end{equation}
The reader is invited to make a comparison with the 
definition of the conventional $S$-matrix of closed systems 
for which $|{\rm out}> = S\, |{\rm in}>$. 
In such systems one can defined a von-Neumann density matrix 
$\rho = |\Psi><\Psi|$ for which (\ref{supscat}) yields 
a factorizable \$-matrix: \$$_{\rm closed} = SS^\dagger$. 
But in open systems
\$ 
is not factorizable, \$~$\ne SS^\dagger$, and unitarity of the theory is 
lost. This has been argued by Hawking to characterize 
generic quantum-gravity theories. Such features may lead to  
effects which can be, in principle, observable~\cite{ehns}.

However, the above intuitive description, may be incomplete. 
The apparent `loss of information' may not be a fundamental problem of 
quantum gravity, given that the 
above arguments pertain strictly to the dynamics of only 
part of the complete
system, that of matter living in the `exterior' of the 
(microscopic) black 
hole horizon. If complete knowledge of the interior spacetimes 
(and the associated dynamics) of such  microscopic singular configurations 
were known, then it would be possible, in principle, to trace down 
the information carried by the relevant internal (`gravitational') 
degrees of freedom
of the black hole. 

String theory~\cite{GSW}, and its modern 
non-perturbative membrane version (D(irichlet) branes)~\cite{branes}, 
provides the first, and probably the only up to date, 
mathematically consistent theory where such issues began to be addressed
in a rather rigorous way. Macroscopic 
Black Holes are viewed as solitonic states
in this approach, with a rich internal dymamical structure, that can account
for a great deal of the information carried by the relevant 
degrees of freedom. In general terms, stringy black holes carry 
an infinity of energy (Planckian) levels, which are thus capable 
of accounting for the enormous entropy associated with such 
(macroscopic) objects,
and thus for the apparent loss of information. An exact 
counting of the 
internal states of black holes 
has been made possible {\it for specific cases} 
~\cite{duality}, using   
D(irichlet) branes~\cite{branes}. 

The underlying duality symmetries 
are discrete gauge symmetries in string theory,
believed to be exact, which help  mapping a strongly coupled 
non-perturbative problem, 
such as the counting of microstates of a black-hole soliton,
to a dual model where perturbative methods (in some sense) can be applied.    
In some models the horizon of the black hole itself is treated as
a $D$-brane. 
However, such exact (non-perturbative) results are only possible for very 
specific string backgrounds, and refer only to macroscopic black holes.
At present there is no exact treatment, in our opinion, 
that allows for generic {\it microscopic} 
black hole configurations, which 
appear in any complete theory of quantum gravity, 
to be treated in a satisfactory way. 
From this point of view, the problem of finding a 
mathematically and physically consistent theory of 
stringy quantum gravity is still unsolved. 
To this problem one should add the above-mentioned 
problem of quantizing
eternally accelerating Universes~\cite{accel}
with global particle horizons
within a critical string theory framework. 

It is our belief that such  problems cannot be treated 
by restricting oneself to equilibrium field theory methods, such as 
studies of 
conformal backgrounds of critical string/membrane theory only, which 
constitute the dominant part of the contemporary string literature. 
In fact, 
there are arguments in the string literature~\cite{myers},
supporting the fact that the process of formation and subsequent evaporation
of a black hole, microscopic or macroscopic, cannot be described
entirely within the framework of 
equilibrium field theories, but necessarily 
involve concepts of {\it quantum decoherence}. In other words, {\it 
pure quantum 
states cannot form black holes}.  Such arguments were derived 
within the above-mentioned modern framework of D(irichlet) branes 
for counting black-hole microstates. In such a framework,
one gives a statistical meaning to the black-hole entropy 
by matching it with a quantity that counts the internal states.
Hence, in the relevant $D$-brane calculations~\cite{dbranebh}, 
the black hole results for the entropy can be matched  
only by considering a {\it mixed} state, including all the degenerate
configurations of the $D$-brane. 
Moreover, 
in those works it was found that, in order to 
calculate the absorption cross section of a scalar mode incident  
on a $D$-brane configuration, one must include a {\it decoherent ensemble} 
of all such 
degenerate $D$-brane configurations.

An additional 
argument in favour of a stochastic treatment
of stringy quantum black holes is provided by 
the fact that 
$D$-brane calculations refer to weak coupling theories, 
which are {\it assumed} to 
be connected with strongly-coupled gravitational field theory solutions 
by means of appropriate duality symmetries. 
Near the high curvature regions of strong gravitational 
fields, the concept of space time, at least as 
we preceive it in the context of low-energy physics, may itself
break down. In this sense the apparent black hole singularities
may be a property only of the truncation of the dynamics to the 
low-energy Einstein action, which certainly does not describe the 
strong gravity region of black hole singularities in string theory.
The latter may well resemble the analogous situation 
in the very Early Universe, where the associated strong gravitational
field is treated by many researchers in a stochastic (non-equilibrium) 
way.

In \cite{emn} we have taken some modest steps to 
formulate mathematically such non-equilibrium processes
in string theory, by adopting the point of view that 
{\it departure from equilibrium}, in the context of string theory, 
is linked to {\it departure from criticality}. 
Non-critical strings~\cite{ddk} appear to be mathematically
consistent in the same way as critical strings are, at least
from a (perturbative) world-sheet point of view. 
The idea of non-critical strings can be summariuzed as follows: 
there are some deformations
in the $\sigma$-model, which cause deviations of the associated 
world-sheet theory  from conformal invariance. 
Such deviations imply the coupling of another world-sheet field, 
the Liouville mode $\phi$, whose presence restores  
conformal invariance~\cite{ddk}.

There are two cases of non-critical strings,
depending on whether the central charge 
of the deformed $\sigma$-model  
is less or bigger 
than  a critical (fixed-point) value (which is $25$ for bosonic strings and
$9$ for supestrings). In the former 
case ({\it subcritical} strings), 
the Liouville mode $\sigma$-model kinetic terms 
appear with space-like
signature, i.e. with the same sign as the 
$\sigma$-model kinetic terms of the space components of the target space time. 
In the second case ({\it supercritical} string), the Liouville
mode has a {\it time-like} signature (i.e. it appears as a ghost field
in the $\sigma$-model). 

In the second case the Liouville mode can be considered as 
a target time variable. In \cite{emn} we have developed an approach,
according to which a
time-like Liouville field is {\it identified} with the 
observable target time. This approach results in a 
{\it non-equilibrium} treatement, which is manifested 
in many ways.
These 
include entropy production in such theories, and in general 
the presence of instabilities. Such features can be seen by 
many researchers 
as unwanted features. 
However, instead of discarding such models, we have adopted the point of view
that such instabilities are physical, and in fact constitute an 
important ingredient of the non-critical string approach to 
Quantum Gravity. As we have argued~\cite{emn}, and review below, 
this non-equilibrium approach is similar in spirit 
with standard non-equilibrium approaches of 
open-field theoretic systems~\cite{ctp}. 
In fact, in such models one can give  
a mathematically 
consistent definition of a superscattering matrix \$, which is
simply given by Liouville correlation functions on the world sheet
among the pertinent vertex operators. Upon the 
identification of the Liouville mode with the target 
time, however, 
such correlators do not yield scattering amplitudes, because they fail
to factorize into products of $SS^\dagger$. Nevertheless,  
as objects in string theory,
are well defined~\cite{emn,emnsmatrix}. 

A natural question arises at this point,
concering the predicting power of such Liouville strings.
Leaving aside the possibility of resolving conceptually important 
formal issues of quantum  
gravity, the big question is whether 
non-critical strings lead to 
low-energy properties which could, in principle, distinguish 
them experimentally
from critical string theories. 
One would hope, and expect, that the situation is analogous to 
that of local field theory, where 
non-equilibrium models of field theory
can be distinguished from equilibrium ones 
due to their stochastic properties. 
For instance, in many of such models there is entropy production
and thermal effects, leading to observable (in principle) violations
of various properties and symmetries of equilibrium theories,
such as Lorentz symmetry.   

It is the point of this article to discuss some physical
properties of a specific non-critical string theory model,
inspired by the modern $D$-brane approach to string theory.
Specifically, we shall consider a toy example of a 
four-dimensional space time punctured by point-like
$D$-particle defects. Such a situation may be viewed as 
a simple case of intersecting branes in string theory. 
The defects are scattered by a (macroscopic) number of 
closed string states, propagating in the space time.
The scattering of the string/$D$-particle is treated in the 
semi-classical {\it impulse} approximation, according to which
the pertinent $\sigma$-model is deformed by 
operators expressing a `sudden' movement of the defect, at a given time 
moment. These deformations 
obey a logarithmic conformal world-sheet 
algebra~\cite{kogan,szabo}, which lie in the border line 
between conformal field theories and general two-dimensional 
field theories, so that they can still be classified 
by means of conformal data.
As such, these $\sigma$-model deformations  
constitute acceptable string theory backgrounds. 
From a world-sheet
renormalization-group view point, 
such operators are marginally relevant. Hence they violate 
(slightly) conformal invariance, 
thereby necessitating Liouville dressing~\cite{ddk}.
The dressing results in induced space time deformations
in the neighbourhood of the recoiling defect, 
whose physical properties we study below in detail. 

In the present article we shall consider the $D$-particle defects as 
being real.
In this sense, our analysis below is related to quantum gravity considerations
only in that we consider quantum metric fluctuations 
about a specific background space time.
In a $\sigma$-model perturbative context this is achieved
by summing up world-sheet topologies (genera). 
The full non-perturbative
problem of considering {\it virtual} space time fluctuations
{\it in vacuo} is not solved, as it lies beyond our 
calculational capability at present.
Nevertheless, as we shall see, the results obtained from the
analysis in this simplified problem, look, at least to the authors,
sufficiently interesting and non-trivial to motivate further studies
along this direction, and also to imply important  thermal properties
that might characterize space-time foam situations of 
quantum gravity~\cite{garay}. 
As we shall see, one 
aspect of our analysis 
will be an effective violation of Lorentz symmetry, precisely due to 
the aforementioned thermal properties, which will manifest 
itself as a non-trivial refractive index. But as we shall argue,
such effects will not be detectable in the near futrure.
What could be detectable, though, is the emission of very high
energy photons from the recoiling $D$-particle defects, 
which could be linked
to some extreme astrophysical phenomena, e.g. those associated with 
ultra-high-energy cosmic rays. Such a possibility 
implies severe phenomenological constraints on the model,
which will be discussed briefly. 

The structure of the article is as follows: in section 2 we 
study some formal aspects of the Liouville (non-critical) string dynamics,
with emphasis on 
demonstrating their non-equilibrium nature.  
In section 3 we review in some detail 
a specific example of non-critical string models, 
in which the non-criticality is a consequence of the scattering
of a $D$-particle defect, embedded in our four-dimensional space time,
with (a macroscopic number of) closed string states in the 
semi-classical impulse approximation. 
In section 4 we study the physical properties of the associated 
metric distortion, which results in the formation of a bubble spacetime,
of finite radius
around the defect.
We pay particular attention to denonstrating 
the consistency of the approach with 
the (lowest order) $\sigma$-model conformal invariance conditions.
This is  
a result of the restoration of conformal invariance by means of
Liouville dressing, under the identification of the 
Liouville field with the target time. 
Specifically, we 
show that the interior of the bubble 
is consistent with the 
effective field theory obtained from strings, with non-trivial 
tachyon-like and antisymmetric tensor fields. However 
the tachyonic mode expresses an instability of the 
distorted spacetime, which is argued to exist also 
in supersymmetric strings.  For the validity of our 
perturbative approach 
it is essential that the string length is a few orders of magnitude 
larger than the Planck length. This 
is  a typical situation encountered
in many $D$-brane-world models.
In section 5 we analyse thermal effects and entropy production, 
and argue on the emission of 
high energy photons from such geometries. 
The entropy production 
is due to the non-trivial matter content of the 
interior of the bubble, and is associated 
with information being carried away by the `recoil' degrees of freedom.
In section 6 we analyse the motion of particles in the interior 
of our unstable space time, and demonstrate capture of particles,
associated with loss of information for an asymptotic observer. 
The latter becomes manifest through entropy 
production, proportional to the area of the unstable space-time bubble.
In section 7 we discuss the appearance of effective 
{\it refractive indices} due to the aforementioned 
non-trivial thermal effects
in the interior of the bubble. 
In section 8 we speculate on a possible 
link of these geometries to the physics of extreme
astrophysical phenomena, in partucular to ultra high energy cosmic rays,
and argue how such phenomena can be used to constrain 
some parameters of our non-critical string theory model.
In this respect we argue that observable phenomena from such models may only come if one has a certain population of $D$-particles
in the Universe, and we use recent astrophysical observations 
to constrain the maximum number of constituent defects in such 
populations.  
In section 9 we make some remarks on 
the important issue of stability of 
these $D$-particle populations, in the modern context of 
non-perturbative string theory. In section 10
we comment briefly on the r\^ole of the embedded recoiling 
$D$-particles on the 
cosmological evolution of a Friedmann-Robertson-Walker type Universe. 
We argue in favour of an eventual stopping of the 
cosmic acceleration, thereby implying the absence of 
cosmological horizons in such models, and hence the possibility 
of a proper definition of asymptotic states. This is linked 
to the fact that our non-critical string model asymptotes
to a critical (equilibrium) string for large times.  
We also demonstrate that such cosmological models may be characterised
by a relaxing-to-zero vacuum energy a l\'a quintessence. 
Finally, concluding remarks are presented in section 11.

\section{Generic Non-Equilibrium Aspects of Liouville Strings}

We commence our analysis with a brief review of the Liouville
dressing procedure for non-critical strings, with the Liouville
mode viewed as a local world-sheet renormalization group
scale~\cite{emn}. Consider a conformal $\sigma$-model, described
by an action $S^*$ on the world-sheet $\Sigma$, which is deformed
by (non conformal) deformations $\int_{\Sigma}g^iV_id^2\sigma$,
with $V_i$ appropriate vertex operators.
\begin{equation}
S_g = S^* + \int_{\Sigma}g^iV_id^2\sigma
\label{sigma}
\end{equation}
The non-conformal nature of the couplings $g^i$ implies that their
(flat)world sheet renormalization group $\beta$-functions,
$\beta^i$, are non-vanishing. The generic structure of such
$\beta$-functions, close to a fixed point, $\{g^i = 0\}$ reads:
\begin{equation}
\beta^i = (h_i - 2)g^i + c^i_{jk}g^jg^k + O(g^3).
\label{fixed}
\end{equation}
where $h_i$ are the appropriate conformal dimensions. 
In the context of Liouville strings, world-sheet gravitational dressing is
required. The ``gravitationally''-dressed couplings, $\lambda^i(g,\phi)$,
which from our point of
view correspond to renormalized couplings in a curved space,
read to $O(g^2)$~\cite{ddk,liouville}:
\begin{equation}
\lambda^i(g,\phi) = g^ie^{\alpha_i\phi} + \frac{\pi}{Q + 
2\alpha_i}c^i_{jk}g^j g^k\phi e^{\alpha_i\phi} + O(g^3), \qquad
Q^2 = \frac{1}{3}(c-c^*)
\label{renorm}
\end{equation}
where $\phi$ is the (world-sheet zero mode) of the Liouville field, 
and $Q^2$ is the central
charge deficit, 
with $c=c[g]$ the (`running') central charge of the deformed
theory~\cite{zam}, and $c^*$ one of its critical values (conformal point)
about which the theory is perturbed by means of the operators
$V^i$. Close to a fixed point $Q^2$ may be considered
as independent of $g$, but this is not true in general. 
Finally, $\alpha_i$ are the gravitational anomalous dimensions:
\begin{equation}
\alpha_i(\alpha_i + Q) = 2 - h_i \qquad {\rm for} \qquad  c \ge c^*
\label{anomcoeff} 
\end{equation} 
Below we shall concentrate exclusively
to the supercritical string case, $Q^2 \ge 0$, which from the
point of view of identifying the Liouville mode with target time,
corresponds to a Minkowskian signature spacetime manifold.

Due to the renormalization (\ref{renorm}), the critical-string
conformal invariance conditions, amounting to the vanishing of
flat-space $\beta$-functions, are now substituted by:
\begin{equation}
{\ddot \lambda}^i + Q{\dot \lambda}^i = -\beta^i(\lambda) + \dots \qquad
{\rm for}~ c \ge c^*.  
\label{neweq} 
\end{equation}  
where the overdot denotes derivative with respect to the Liouville
mode $\phi$, and the 
$\dots $ denote higher-order terms, quadratic in ${\dot \lambda^i}$,
${\cal O}\left(({\dot \lambda}^i)^2\right)$.  
As we shall argue later, such terms 
can either be removed by field redefinitions,
or alternatively are negligible if one works in the neighbourhood of 
a world-sheet renormalization-group fixed point, which is the case we shall
consider in this work. 
The
notation $\beta^i(\lambda)$ denotes flat-world-sheet
$\beta$-functions but with the formal substitution $g^i
\rightarrow \lambda^i(g,\phi)$. Note the minus sign in front of the
flat world sheet $\beta$-functions $\beta^i$ in (\ref{neweq}),
which is characteristic of the supercriticality of the
string~\cite{ddk,liouville}. 
Notice that upon the identification of the Liouville mode 
$\phi$ with the target time $t$ 
the overdot denotes temporal derivative.

In \cite{emn} we have treated the Liouville mode as a local
(covariant) world-sheet renormalization-group  scale. 
To justify formally this interpretation, one may write
\begin{eqnarray} \label{liouvareatime}
&~&\phi = -\frac{2}{\alpha}\tau\, , \qquad \tau \equiv  
-\frac{1}{2}{\rm log}A\, , \qquad 
A = \int _{\Sigma} d^2\sigma \sqrt{\gamma}=
\int _{\Sigma} d^2\sigma \sqrt{{\widehat \gamma}}e^{\alpha \, \phi}\, , 
\nonumber \\  
&~& \alpha=-\frac{Q}{2} + \frac{1}{2}\sqrt{Q^2 + 8}\, , \nonumber \\
\end{eqnarray}
where $\gamma$ is a world-sheet metric, and ${\widehat \gamma}$ 
is a fiducial metric, obtained 
after the conformal gauge choice in terms of  
the Liouville mode $\phi$~\cite{ddk}. 
We thus observe that the Liouville mode is associated with the 
logarithm of the world-sheet area $A$. 

Using (\ref{liouvareatime}), we can re-write (\ref{renorm}) 
in a standard ``flat-world-sheet''
renormalization-group form~\cite{emn,schmid}: 
\begin{eqnarray}
&~& \frac{d}{d\tau}\lambda^i = ({\tilde h}_i - 2)\lambda^i + \pi {\tilde c}^i_{jk}\lambda^j\lambda^k +  \dots\, , \nonumber \\
&~& {\tilde h}_i - 2 =-\frac{2}{\alpha}\alpha_i\, , 
\qquad {\tilde c}^i_{jk}=-\frac{2}{\alpha (Q + 2\alpha_i)}c^i_{jk}\, .  
\label{flatrenorm}
\end{eqnarray}
which justifies formally the identification~\cite{emn} of the 
Liouville mode with a local renormalization-group 
scale on the world sheet. 
It also implies that the point $\phi \to  \infty$ is an 
infrared fixed point of the flow, in which case the 
world-sheet area diverges $|A| \to \infty$.

A highly non-trivial feature of the $\beta^i$ functions is the
fact that they are expressed as gradient flows in theory
space~\cite{zam,osborn}, i.e. there exists a `flow' function
${\cal F} [g]$ such that
\begin{equation}
\label{flow} \beta^i = {\cal G}^{ij}\frac{\delta {\cal F}[g]}{\delta g^j}
\end{equation}
where ${\cal G}^{ij}$ is
the inverse of the Zamolodchikov metric in theory
space~\cite{zam}, 
which is given by appropriate two-point
correlation functions between vertex operators $V^i$,
${\cal G}_{ij} \sim <V_i V_j>$. 
In the case
of stringy $\sigma$-models the flow function ${\cal F}$ may be
identified~\cite{osborn} with the running central charge deficit
$Q^2[g]$. 

An important comment we would like to make concerns the
possibility of deriving the set of equations (\ref{neweq}) from a
target space action. This issue has been discussed in the
affirmative in \cite{emn}, where it was shown that the set of
equations (\ref{neweq}) satisfies the Helmholtz conditions for the
existence of an action in the `space of couplings' $\{ g^i \}$ of
the non-critical string. The property (\ref{flow}) is crucial to
this effect. Upon the identification of target time with the
Liouville mode~\cite{emn} this action becomes identical with the
target space action describing the off-shell dynamics of the
Liouville string. We should stress the fact that the action is off
shell, in the sense that the on-shell conditions correspond to the
vanishing of the $\beta$-functions $\beta^i$,
while  in our case $\beta^i
\ne 0$. 

However, the restoration of conformal invariance by the Liouville 
mode implies that in an enlarged target space time,
with coordinates $(\phi, X^0, X^i)$ the resulting 
$\sigma$-model will be conformal, for which 
one would have the normal conformal invariance conditions~\cite{GSW}. 
This means that the set of equations (\ref{neweq}) can be cast 
in a conventional form, amounting to the vanishing of $\beta$ functions
of a $\sigma$-model, but in this enlarged space:
\begin{equation}\label{odh} 
 {\tilde \beta}^{(D+1)}(\lambda) =0 
\end{equation} 
where $D$ is the target-space dimensionality of the $\sigma$-model
before Liouville dressing, and 
now there are Liouville components as well in 
appropriate tensorial coordinates.

For fields of the string multiplet, it can be checked explicitly 
that (\ref{odh}) and (\ref{neweq}) (in $D$-dimensions) 
are equivalent~\cite{schmid}. 
For completeness, we shall demonstrate this by considering 
explicitly the dilaton $\Phi$, graviton $G_{\mu\nu}$ 
and antisymmetric tensor fields $B_{\mu\nu}$.
We shall not consider explicitly the tachyon field, although its
inclusion is straightforward and does not modify the results.
This is because, as we shall see 
in the next section, in our non-critical string model the tachyon-like
mode expresses only the instability of the spacetime metric, 
and thus in principle it should also appear in supersymmetric
strings, which do not have normal (flat-space-time spectrum) tachyons. 

To ${\cal O}(\alpha ')$, the appropriate $\sigma$-model $\beta$-functions
for a $D$-dimensional target spacetime, 
parametrized by coordinates $X^\mu$, $\mu=0, 1, \dots D-1$, 
read~\cite{GSW}: 
\begin{eqnarray} 
{\widehat \beta}^{\Phi (D)} & = & \beta^{\Phi (D)} - \frac{1}{4}
G^{\mu\nu}\beta^{G(D)}_{\mu\nu}=\frac{1}{6}\left(C^{(D)} - 26\right)\, ,
\nonumber \\
C^{(D)} &=& D -\frac{3}{2}\alpha ' \left(R - \frac{1}{12}
H_{\mu\nu\rho}H^{\mu\nu\rho} - 4 (\nabla \Phi)^2 + 4 \nabla ^2 \Phi \right)\, ,
\nonumber \\
\beta^{G(D)}_{\mu\nu} & = & \alpha ' \left(R_{\mu\nu} + 2 \nabla_\mu \nabla_\nu \Phi -
\frac{1}{4}H_{\mu\sigma\rho}{H_{\nu}}^{\sigma\rho} \right)\, , \nonumber \\
\beta^{B(D)}_{\mu\nu} & = & \alpha ' \left(-\frac{1}{2}\nabla_\rho {H^\rho}_{\mu\nu}
+ {H^\rho}_{\mu\nu}\nabla_\rho \Phi \right)\, .
\label{betafnct}
\end{eqnarray} 
where $H_{\mu\nu\rho}=3\nabla_{[\mu}B_{\nu\rho]}$ is the 
antisymmetric tensor field-strength, on which the 
$\beta$-functions depend, as dictated by an appropriate 
Abelian Gauge symmetry~\cite{GSW}.

To demonstrate that such $\beta$-functions yield
equations of the form (\ref{neweq}), when they are reduced
to a target-space manifold with one lower dimension, 
we separate from the expressions (\ref{betafnct}) 
a Liouville component. 
We first note that there is a special normalization 
of the $\sigma$-model kinetic term of the Liouville field $\phi$ 
for which (\ref{anomcoeff}) is valid, which implies
that the enlarged spacetime metric is of 
``Robertson-Walker'' form with respect to $\phi$, i.e.: 
\begin{equation}
ds^2 = -d\phi^2 + G_{\mu\nu}(\phi, X^\mu)dX^\mu\, dX^\nu\, , \quad \mu, \nu = 0, 1, \dots D-1
\label{rwliouv}
\end{equation} 
where the Minkowski signature of the 
Liouville term is due to the assumed supercriticality 
of the non-critical string~\cite{ddk,aben89}. 
This implies that for graviton and antisymmetric
tensor $\beta$-functions one has: 
\begin{equation}
{\tilde \beta}^G_{\phi\phi}={\tilde \beta}^{G,B}_{\phi\mu}=0
\label{constr}
\end{equation} 
which are viewed as additional constraints. 
However, from the point of view of the enlarged space time such 
constraints can be easily achieved by an appropriate
general coordinate transformation, which from our point of view
is a renormalization-scheme choice.

We find it convenient to shift the dilaton~\cite{schmid}:
\begin{equation} 
   \Phi \to \varphi = 2\Phi - {\rm log}\sqrt{G} 
\end{equation} 
In this case we may write (\ref{betafnct}) 
as follows (to keep consistency with the previous notation
we have denoted the $\beta$-functions in the enlarged spacetime 
$(\phi, X^\mu)$ by ${\tilde \beta}$):
\begin{eqnarray} 
&~& 0 = C^{(D+1)} - 26 = C^{(D)} - 25 - 3G^{\phi\phi}\left({\ddot \varphi} - 
({\dot \varphi})^2 \right), \nonumber \\
&~& 0 = {\tilde \beta}^{G}_{\phi\phi} = 2{\ddot \varphi} -
\frac{1}{2}G^{\mu\kappa}G^{\nu\lambda}\left({\dot G}_{\mu\nu}{\dot G}_{\kappa\lambda} + {\dot B}_{\mu\nu}{\dot B}_{\kappa\lambda}\right), \nonumber \\
&~& 0 = {\tilde \beta}^{G}_{\mu\nu} = \beta^{G(D)}_{\mu\nu}
-G^{\phi\phi}\left({\ddot G}_{\mu\nu} -{\dot \varphi}{\dot G}_{\mu\nu}
- G^{\kappa\lambda}[{\dot G}_{\mu\kappa}{\dot G}_{\nu\lambda} -
{\dot B}_{\mu\kappa}{\dot B}_{\nu\lambda}]\right), \nonumber \\
&~& 0 = {\tilde \beta}^{B}_{\mu\nu} 
= \beta^{(D)}_{\mu\nu} - G^{\phi\phi}\left({\ddot B}_{\mu\nu}
- {\dot \varphi}{\dot B}_{\mu\nu} - 2G^{\kappa\lambda}
{\dot G}_{\kappa[\mu}{\dot B}_{\nu]\lambda} \right)\, .
\label{dimred}
\end{eqnarray} 
where the overdot denotes total Liouville scale derivative. 
In our interpretation of the Liouville field as a (local) 
renormalization scale~\cite{emn} this 
is equivalent to a total world-sheet renormalization-group derivative. 

In Liouville strings~\cite{ddk}, 
the dilaton $\Phi$, as being coupled to 
the world-sheet curvature, receives 
contributions from the Liouville mode $\phi$ which are linear.
In this sense one may split the dilaton field 
in $\phi$-dependent parts and $X^\mu$ dependent parts
\begin{equation}
\Phi (\phi, X^\mu) = -\frac{1}{2}Q\phi + {\tilde \Phi}(X^\mu)
\end{equation}
where $Q^2 = \frac{1}{3}\left(C^{(D)} - 25\right)$ 
is the central charge deficit, and the normalization 
of the term linear in $\phi$ is dictated by the analysis 
of \cite{ddk}, in which the Liouville mode has a canonical
$\sigma$-model kinetic term.
This implies
that $\varphi$ is such that: 
\begin{equation}
{\dot \varphi} = -Q +{\cal O}(\sqrt{G}G^{\mu\nu}{\dot G}_{\mu\nu})\, , 
\label{varphifirst}\end{equation}
Therefore, 
the renormalization-group invariance of the central charge
$C^{(D)}$, implies ${\dot Q}=0$. 
Note that, in the context of the equations 
(\ref{dimred}), the terms in ${\dot \varphi}$ 
proportional to ${\dot G}_{\mu\nu}$, will yield
terms quadratic in Liouville derivatives of fields. 

Upon our interpretation of the Liouville field as a (local) 
renormalization scale~\cite{emn} 
terms quadratic in 
the Liouville derivatives of fields, 
i.e. terms of order 
${\cal O}\left({\dot G}{\dot B},{\dot G}{\dot G},{\dot B}{\dot B}\right)$
become quadratic in appropriate $\beta$-functions. 
Such quadratic terms may be removed by appropriate 
field redefinitions~\cite{remark}, provided the gradient flow 
property (\ref{flow}) is valid, which can be shown to be 
true
for the Liouville local renormalization-group world-sheet 
scale~\cite{emn,osborn}.   
Alternatively, one may ignore such quadratic terms in Liouville derivatives
of fields by working in the neighbourhood of a renormalization-group 
fixed point. Such terms are of higher order in a weak-field/
$\sigma$-model-coupling expansion, and thus can be safely neglected
if one stays close to a fixed point.
This is the case of the specific example of recoiling $D$-particles,
where one has only marginal non-criticality for slow-moving heavy 
$D$-particles, as we shall see in the next section. 
Ignoring such higher-order terms, 
therefore, and taking into account 
world-sheet renormalizability, one obtains 
\begin{equation} 
{\ddot \varphi} = {\dot Q}+ \dots =0 
\label{varphisecond}\end{equation} 
where the 
$\dots $ denote the neglected 
(higher-order) terms.

Taking into account that $G_{\phi\phi}=-1$ 
for supercritical strings~\cite{aben89} (c.f. (\ref{rwliouv})), 
we observe 
that, 
as a result of (\ref{varphifirst}), (\ref{varphisecond}),  
the first two of the equations (\ref{dimred})
are satisfied automatically (up to 
removable terms quadratic in Liouville derivatives of fields).
The first of these equations 
is the dilaton equation, which thus  
becomes equivalent to the definition of $Q^2$, and therefore 
acquires a trivial 
content in this context.  
Notice also that the second of these equations
is due 
to the constraints (\ref{constr}), which should be taken into account 
together with the set of equations (\ref{dimred}).
It can be shown~\cite{schmid} that the rest of these constraints
do not impose further restrictions, and thus can be ignored, 
at least close to a fixed point, where the constraints 
can be solved for arbitrary $G_{\mu\nu}, B_{\mu\nu}$ fields. 
The rest 
of the equations 
(\ref{dimred}) then, for graviton and 
antisymmetric tensor fields, 
reduce to (\ref{neweq}), up to irrelevant terms quadratic 
in Liouville derivatives of fields.
This completes our proof
for the case of interest.

What we have shown above is that the Liouville equations 
(\ref{neweq}) can be obtained from 
a set of conventional $\beta$-function equations (\ref{odh}) 
if one goes to a $\sigma$-model with one more target-space 
dimension, the extra
dimension being provided by the Liouville field. 
We now stress that 
the identification of the Liouville mode with the target time
$X^0$, which distinguishes the approach of \cite{emn,recoil}
from the standard Liouville approach described above
in which $\phi$ was an independent 
mode, 
will be made in expressions of the form 
(\ref{odh}), pertaining to the enlarged $(D+1)$-dimensional 
spacetime $(\phi, X^\mu)$. 
Then, one should look
for consistent solutions of the resulting equations
in the $D$-dimensional submanifold $(\phi=X^0, X^i)$. 
In this sense, the target-space dimensionality 
remains $D$, but the resulting string will be characterized 
by the Liouville equations 
(\ref{neweq}), supplemented by the constraint of the 
identification $\phi=X^0$, and will have  
a non zero central charge deficit $Q^2$, appearing as target-space
vacuum energy~\cite{GSW},  
in contrast to 
the case of 
treating the Liouville mode as an independent coordinate.

To put it in other words, 
one starts from a critical $\sigma$-model, perturbs it
by the recoil deformation, induces non-criticality, but
instead of using an extra Liouville $\sigma$-model field, one 
uses the existing time coordinate as a Liouville mode, i.e 
one invokes a readjustment  
of the time dependence of the 
various background fields (a sort of back reaction), in order to restore the 
broken conformal invariance.  
It is a non-trivial fact that there are consistent solutions to 
the resulting equations, and this is the topic of the present work, 
namely we shall consider a specific model of non-critical
strings, in which we shall identify the time with the 
Liouville mode, and we shall present consistent solutions 
of (\ref{odh}), under the constraint $\phi=X^0$,  
to ${\cal O}(\alpha ')$ in the respective $\sigma$-model. 
This is done in the next two sections.

\begin{figure}
\epsfxsize=2.in
\bigskip
\centerline{\epsffile{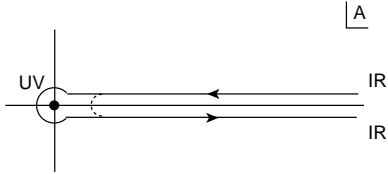}}
\vspace{0.2in}
\caption{Contour
of integration for a proper definition of the path 
integration for the Liouville field, where
the quantity $A$ denotes the (complex) world-sheet area. 
Upon the interpretation of the 
Liouville field with target time, this curve
resembles closed-time-paths in non-equilibrium field theories.} 
\label{dollar}
\end{figure}

We would now like to comment briefly 
on the breakdown of the interpretation 
of world-sheet correlators in Liouville strings 
as unitary scattering amplitudes in the target space time.
This is one of the most important features 
which makes Liouville strings non-equilibrium string theories~\cite{emn}. 
Below we recall briefly our formal arguments~\cite{emn} supporting the point
of view that it is impossible to define a unitary $S$ matrix in Liouville
strings, with the only well-defined object being the \$ matrix.  This
follows from the fact that an $N$-point world-sheet correlation function of
vertex operators in Liouville strings, ${\cal F}_N \equiv <V_{i_1} \dots
V_{i_N}>$, where $<\dots>$ signifies a world-sheet expectation value in
the standard Polyakov treatment, transforms under infinitesimal Weyl
shifts of the world-sheet metric $\gamma$ in the following
way:
\begin{equation}
\delta_{\rm weyl} {\cal F}_N \, = \, 
\left[ \delta_0
+ {\cal O}\left(\frac{s}{A}\right)\right]{\cal F}_N
\label{weyl}
\end{equation}
where the standard part $\delta_0$ of the variation
involves a sum over 
the conformal dimensions $h_i$ of the operators $V_i$ that is independent
of the world-sheet area $A$ (whose logarithm is the
world-sheet zero mode of the Liouville field). 
The quantity $s$ is the sum of 
the gravitational anomalous dimensions~\cite{ddk}:
\begin{equation}
s=-\sum_{i=1}^{N}\frac{\alpha_i}{\alpha} - \frac{Q}{\alpha}~, \quad 
~\alpha_i=-\frac{Q}{2}+ \frac{1}{2}\sqrt{Q^2 + 4(2 - h_i)}, \quad 
~\alpha=-\frac{Q}{2}+ \frac{1}{2}\sqrt{Q^2 + 8}
\label{gad}
\end{equation}
where $Q^2$ is the central-charge deficit, that is non-zero for
non-critical string. 

From (\ref{weyl}), and upon the identification of the logarithm of the
world-sheet area ${\rm ln}A$ with the target time $t$~\cite{emn}, one
observes that such a correlator cannot represent a unitary $S$-matrix
element in target space, as a result of the explicit $A-$ (i.e. time $t-$)
dependence. Nevertheless, according to the analysis
in~\cite{emn},
it is
clear that one can construct a well-defined quantity free from
such world-sheet area ambiguities.  However, this is a \$ matrix that is
not factorizable into a product of $S$ and $S^\dagger$. This results from
the way one performs the world-sheet path integration over the Liouville
mode $\int D\phi (\dots )$ in a first-quantized approach to string theory. 
This integration can be done via a steepest-descent
method~\cite{emn,kogan2}, along the curve in
Fig.~\ref{dollar}, which is reminiscent of the closed time-like
path used in non-equilibrium field theories~\cite{ctp}. Close to the
ultraviolet fixed point on the world sheet, i.e., in the limit when $A \to
0$, there are divergences. These are responsible for the lack of
factorization of the \$ matrix in this case. 

The definition of the
Liouville path integral over such a curve is responsible, in this
formalism, for a density-matrix rather than a
wave-function interpretation of the world-sheet partition function of the
Liouville string, in full agreement with its non-equilibrium
(open-system) nature. Indeed, the effective action $\Gamma [g^i]$ in 
the space of couplings $g^i$ of the closed string theory 
under consideration is connected to the Liouville path integral
as follows:
\begin{equation}
e^{-\Gamma[g^i]} = \int D\phi DX^i e^{-\int _{\Sigma}d^2 \sigma 
{\cal L}(g^i,\phi=t, 
X^i)}
\end{equation} 
where ${\cal L}$ is the Liouville-dressed world-sheet Lagrangian 
of the deformed
$\sigma$-model (\ref{sigma}) after the identification
of the Liouville mode $\phi$ with the target time $t$. 
The extension of the Liouville zero mode integration over the contour
of figure~\ref{dollar} implies that the time integration in $\Gamma [g^i]
=\int dt L[g^i(t), {\dot g}^i(t), t]$
will result 
in products $\Psi (g^i, t)\Psi^*(g^i, t)$ at each time moment, 
with $\Psi (g^i, t)$ the wave function  
at a fixed slice of the Liouville time, being defined 
along, say, the upper side curve of figure \ref{dollar},
with its complex conjugate $\Psi^*(g^i,t)$ being defined on the lower side 
curve. 
This yield a probability density interpretation of the Liouville
$\sigma$-model partition function in the space of couplings $g^i$. 
In the case of recoil we are considering in the next section,
the couplings $g^i$ correspond to collective coordinates 
of the $D$-particles.

This impossibility of defining a consistent unitary $S$-matrix element in
Liouville string, but only a \$ matrix, provides an argument that
non-critical string theory may be the key~\cite{emnsmatrix} to a
resolution of the accelerating Universe puzzle, where the presence of a
cosmological (particle) horizon seems to prevent a consistent definition
of a unitary $S$ matrix~\cite{accel}, or in general to 
quantize consistently models of quantum gravity plagued 
by microscopic singular fluctuations with event horizons.

Thus, in Liouville strings, 
the state of the observable system may be characterized by a density matrix
$\rho$. The full density matrix of string theory might in principle be 
a pure-state
density matrix $\rho =|\Psi > < \Psi |$. However, in our
effective non-critical string 
picture $|\Psi > = |\psi, \tilde\psi > $, where $|\psi >$ denotes a state
of the
observable system and $|\tilde\psi >$ the unobserved degrees of freedom that are
integrated out in the renormalization-group approach. The reduced observable
density matrix
${\tilde \rho} \equiv \int d\tilde\psi \,\rho = \int d \tilde\psi | \psi, \tilde\psi > <
\tilde\psi, \psi|$ will in general be mixed as a result of entanglement
with the
unobserved degrees of freedom $|\tilde\psi >$, e.g., those that disappear across
the macroscopic event horizon in an accelerating Universe, or across a
microscopic event horizon in a model of space-time foam.

The renormalization-group equation for the reduced 
density matrix ${\tilde \rho}$ is easy
to derive and can be cast in the form
\begin{equation}
{\dot {\tilde \rho}} = i[{\tilde \rho}, H] +  \nd{\delta H}{\tilde \rho} 
\label{one}
\end{equation} 
with an explicit form for the
non-Hamiltonian operator $\nd{\delta H}$ in terms of the non-conformal
field couplings $g^i$~\cite{emn}:
\begin{equation}
\nd{\delta H}{\tilde \rho} = \beta^i {\cal G}_{ij}[g^j, {\tilde \rho}]
\label{three}
\end{equation} 
where the $\beta^i$ are the non-trivial 
world-sheet renormalization-group functions of the
couplings/fields $g^i$.
We remind the reader that a consistent quantization of the 
couplings $g^i$, which allows one to view them as quantum-mechanical
operators as above, is provided by a summation over world-sheet
genera~\cite{emn,szabo}. 
Notice that an equation of the
form (\ref{one}) is familiar from the theory of quantum-mechanical systems in
interaction with an environment~\cite{ehns}, 
that is provided in our case by the modes
$|\tilde\psi >$ that disappear across the event horizons.

A final comment we would like to make
concerns entropy growth, which is another generic non-equilibrium feature
of non-critical strings~\cite{emn}. 
Indeed, whenever one has entanglement with an environment over
which one integrates, as in the non-critical string
formalism, one  necessarily encounters entropy  growth. 
Within our Liouville 
renormalization-group formalism, there is a simple
formula for the rate of growth of the entropy ${\cal S}$~\cite{emn}:
\begin{equation}
{\dot {\cal S}} \propto  {\cal G}_{ij} \beta^i \beta^j 
\label{fourb}
\end{equation} 
which is positive semi-definite. If any of the renormalization 
functions $\beta^i
\not= 0$, entropy will necessarily grow. Notice that in view of 
the Zamolodchikov's $c$-theorem~\cite{zam}, 
the rate of change of the entropy ${\cal S}$ is proportional 
to the rate of change (along a world-sheet renormalization-group trajectory)
of the effective (`running') central charge deficit. This is in accordance
with the fact that the deficit counts effective degrees of freedom 
of the relevant $\sigma$-model~\cite{zam,kutasov}. 
Thus, in models with space-time boundaries,
there is a natural change in the available degrees of freedom as time 
evolves, given that states may cross these boundaries and become 
unobservable from an asymptotic observer's point of view.

We are ready now to demonstrate the above features 
by looking at a specific example
of induced non-criticality in string theory.
As we shall see, this specific example 
has the virtue of allowing us to approach these non-equilibrium aspects 
of Liouville strings in a 
more conventional way i.e. using techniques which lie closer
to open-system field theory concepts.

\section{A Specific Non-Critical String Model: \\
Recoil-Fluctuating $D$-particles Embedded in the Universe}

\begin{figure}
\epsfxsize=2.5in
\bigskip
\centerline{\epsffile{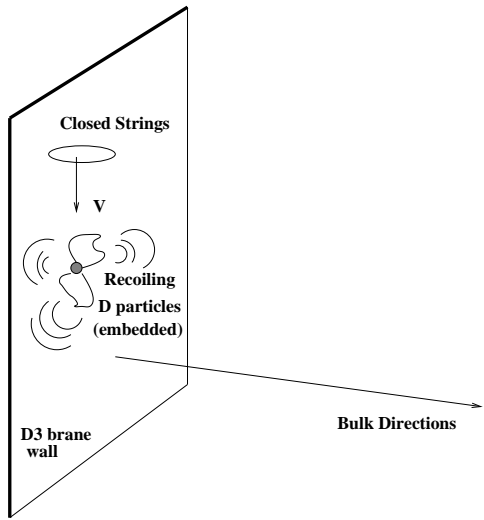}}
\vspace{0.2in}
\caption{Our world viewed as a D3 brane domain wall, embedded
in a higher-dimensional bulk space time. The D3 brane 
is punctured by $D$-particle defects. This setup is a simple
case of intersecting  branes. Recoil fluctuations of the 
$D$-particles due to scattering with closed strings propagating on
the D3 brane distort the spacetime surrounding the defect, and cause 
local curvature. }
\label{branefig}
\end{figure}

We consider for definiteness the situation depicted in figure \ref{branefig}.
Our four-dimensional world is viewed as a three-brane domain wall, punctured 
by $D$-particle defects. From the modern non-perturbative string-theory
perspective, this is an example of intersecting brane configurations.
On the four-dimensional world we have propagating closed  
strings.  When a (macroscopic) number of them strikes a $D$-particle, 
the latter `recoils'. The recoil is treated in the impulse approximation,
which proves sufficient for the case of heavy $D$-particle defects, and weakly
coupled string theories, to which we restrict ourselves for the purposes
of this work. Throughout our approach we shall ignore 
bulk fluctuations of the D3 brane worlds, which will be assumed 
rigid. Such assumptions are compatible with the 
orbifold constructions of \cite{RS}, 
which have attracted enormous attention recently, in view of their
capability of providing a resolution to 
the mass hierarchy problem on the observable world.  
We note that fluctuations of the D3 brane world along the bulk directions
have been considered, in the present context, in \cite{leonta}.

From a $\sigma$-model viewpoint, the `recoil' of the $D$-particle
is described by adding the following deformation to the $\sigma$-model
Lagrangian, 
describing fluctuations of the $D$-brane defect~\cite{kogan,szabo}: 
\begin{equation}\label{recoilop}
{\cal V}_i = \int _{\partial \Sigma} (\epsilon y_i + u_i X^0 )\Theta_\epsilon 
(X^0) \partial_n X^i 
\end{equation} 
where $\partial \Sigma$ denotes the world-sheet boundary, $X^0,X^i$,
$i=1,2,3$  
are four-dimensional 
$\sigma$-model fields, whose zero modes are the space-time coordinates
($X^0$ is the time),
$y_i$ describes the initial location of the $D$-particle defect in space, 
$u_i$ is the recoil velocity in a Galilean approximation due to the 
heaviness of the defects, 
and $\partial_n$ denotes the normal derivative 
on the world-sheet boundary. The deformation (\ref{recoilop})
describes recoil fluctuations of the $D$-particle 
in a comoving frame. 
It can be proven rigorously within the $\sigma$-model framework~\cite{szabo}
that energy and momentum conservation is respected by the 
recoil operators (\ref{recoilop}), and hence the velocity 
$u_i = g_s\frac{\Delta p^i}{M_s}$, where $\Delta p^i$ 
is the momentum transfer during the scattering, 
$M_s$ is the string scale,
$g_s < 1$ is the string coupling, and $M_s/g_s$ is the 
mass of the $D$-particle. 
The operators $\Theta_\epsilon (X^0)$ describe
the impulse approximation, and are regularized Heaviside 
operators~\cite{kogan}:
\begin{equation} 
\Theta_\epsilon (X^0) = \frac{1}{i}\int \frac{d\omega}{\omega -i\epsilon}
e^{i\omega\, X^0}\, \qquad \epsilon \to 0^+ 
\label{regultheta}
\end{equation}
The time coordinate $X^0$ obeys Neumann boundary conditions,
while the coordinates $X^i, \, i=1,\dots 3$ obey 
Dirichlet boundary conditions~\cite{branes}. In 
the limiting case $\epsilon \to 0^+$ only the 
second part of the operator (\ref{recoilop}), proportional to the 
$\sigma$-model coupling $u_i$ is dominant. 

The impulse approximation to recoil may be thought of as the result of 
the incidence of a macroscopic number of closed string states 
on the defect. The impulse approximation is 
semi-classical, which, however, proves sufficient 
for a treatment of heavy $D$-particles, in the context of weakly
coupled strings with couplings $g_s \ll 1$, we are dealing with here.

By studying the appropriate operator products 
of the operators $(\ref{recoilop})$ 
with the world-sheet stress energy tensor, 
it can be shown~\cite{kogan} that the operators are relevant from 
a world-sheet renormalization-group point of view, with anomalous 
dimension
\begin{equation}\label{anomd}
\Delta = -\frac{\epsilon^2}{2}.
\end{equation}
As such, they cause departure from conformal invariance
of the  deformed $\sigma$-model, and thus the need for Liouville 
dressing~\cite{ddk}. 

Before doing so, we remark that the deformations (\ref{recoilop})
obey a logarithmic conformal algebra on the world sheet~\cite{lcft},
provided one has the following relation between the regulating parameter
$\epsilon^2$ and the size $L$ of the world-sheet disc, where the analysis
is performed~\cite{kogan}:
\begin{equation}\label{edef}
\epsilon^2 \sim {\rm log}\left(L/a\right)^2 
\end{equation} 
In the above formula, 
the (omitted) constant proportionality coefficients are of order one, 
$L$ is the size of the world sheet,
and $a$ is an ultraviolet world-sheet cut-off.
The logarithmic pair is provided in this case by 
splitting the operator (\ref{recoilop}) into position and momentum parts:
\begin{equation}
C (z)=\epsilon y_i \Theta_\epsilon (X^0(z))\partial_n X^i(z),
\qquad 
D (z) = \epsilon u_iX^0 (z)\Theta_\epsilon (X^0(z))\partial_n X^i(z),
\end{equation}
with the following operator product expansion on the world-sheet 
boundary~\cite{kogan,szabo}
\begin{eqnarray} 
&~& <C(z) C(w)> \sim {\cal O}(\epsilon^2)\, \to 0\, \nonumber \\
&~& <C(z) D(w) \sim \frac{1}{|z-w|^2}\, \nonumber \\
&~& <D(z) D(w)> \sim \frac{{\rm log}\left(\frac{|z-w|}{L}\right)}{|z-w|^2}
\end{eqnarray}
with $ z \to w$. These are the relations
of logarithmic conformal algebras~\cite{lcft},
which lie in the border line between conformal field theories
and general (renormalizable) two-dimensional field theories.
The presence of the logarithmic algebra is intimately
connected with the fact that the impulse of the $D$-particle defect
implies a sudden change in the background of the $\sigma$-model, and as such
it cannot be described by a conformal field theory.
Nevertheless, the existence of a logarithmic algebra, 
which still allows the classification  
of the appropriate deformations by means of conformal blocks and other
conformal field theory data, implies that such deformed 
$\sigma$-models constitute acceptable backgrounds of string theory.

In the Liouville context, by viewing~\cite{emn} the 
Liouville field as a local renormalization-group  scale on the 
world-sheet, as appropriate for curved-space renormalization~\cite{osborn}, 
the right-hand-side of (\ref{edef}) may be 
connected to 
the zero mode of the Liouville field. As such, we observe from 
(\ref{anomd}) that one encounters a situation
with running anomalous dimensions for the recoil operators (\ref{recoilop}).
The asymptotic point $\epsilon \to 0^+$ corresponds to a 
conformal theory.
By performing the contour integral of the $\Theta_\epsilon (X^0)$ 
we obtain the asymptotic behaviour ($X^0 > 0$) 
$\Theta_\epsilon (X^0) \sim e^{-\epsilon X^0}$,  from which 
we may conclude that 
\begin{equation}\label{time} 
\frac{1}{\epsilon} \sim X^0 \equiv t \gg 0
\end{equation}
This implies that asymptotically in time our non-critical 
(non-equilibrium) string theory of recoiling $D$-particles
relaxes to a conformal (equilibrium) theory. We shall
come back to this point in section 10, when we discuss
possible cosmological implications of this property.

Let us now discuss the physical implications of such a Liouville dressing
procedure in the context of the above model. 
For instructive purposes let us first study the 
simplified case in which the $D$-particle 
defect is recoiling along one direction, say $z$, in space. 
From a modern view point the $D$-particle defect is viewed as a 
zero-space universe (domain `wall'), moving in time $X^0$, and the 
$z$ direction of recoil is viewed as a `bulk' dimension~\cite{leonta}.
The situation will be easily generalized to three spatial `bulk' dimensions,
as we shall see later on. 
For reasons of formal convergence of world-sheet path integrals
we also assume the time $X^0$ to be Euclidean~\cite{kogan,leonta}. 
As we shall see, the Liouville 
formalism generates automatically a Minkowskian signature,
even in such Euclidean models~\cite{emn,kogan,leonta}. 

To determine the effect of
Liouville dressing on this space-time geometry, 
we write the boundary recoil deformations as  bulk
world-sheet deformations~\footnote{Equivalently, one can dress
the boundary non-critical deformations (\ref{recoilop}) directly,
using methods of boundary non-critical strings.
The formal equivalence of the two approaches 
can be established easily~\cite{gravanis}.}:
\begin{equation}
\int _{\partial \Sigma} {u}_{z} X^0\Theta_\epsilon (X^0)
\partial_n z =
\int _\Sigma \partial_\alpha \left({u}_{z} X^0\Theta_\epsilon (X^0)
\partial ^\alpha z \right)
\label{a1}
\end{equation}
The bulk world-sheet Liouville-dressed operators read~\cite{ddk}: 
\begin{equation}
{\cal V}_{z,L} = \int_{\Sigma} 
e^{\alpha_{0z}\,\phi}\partial_\alpha 
\left(u_{z} X^0\Theta_\epsilon (X^0)
\partial ^\alpha z \right) 
\end{equation}
where $\phi$ is the Liouville mode, and $\alpha_{0z}$
is the world-sheet `gravitational' anomalous dimension~\cite{ddk}:
\begin{equation} 
\alpha_{0z}=-\frac{Q_b}{2} +
\sqrt{\frac{Q_b^2}{4} + \frac{\epsilon^2}{2} }
\label{anom}
\end{equation}
where $Q_b$ is the central-charge deficit of the bulk world-sheet
theory. In the recoil problem at hand, 
the deviation from conformal invariance
is due exclusively to the recoil deformations (\ref{recoilop}).
In this case one can estimate~\cite{kanti98}: 
\begin{equation}
Q_b^2 \sim 
\epsilon^2 u^2_z  > 0
\label{centralcharge}
\end{equation} 
for weak deformations $u_{z} \ll 1$, and hence one 
encounters a {\it supercritical} Liouville theory. 
However, due to the smallness of $\epsilon, u_z$ 
one stays in the neighbourhood of a renormalization-group fixed
point, and hence a perturbative analysis, along the lines
discussed in the previous section, is valid. 
The supercriticality 
implies a {\it Minkowskian signature} Liouville-field kinetic term
in the respective $\sigma$-model~\cite{aben89}, which prompts one
to interpret the Liouville field as a time-like target-space
field. At present we shall
treat the Liouville mode  as a {\it second} time
coordinate~\cite{emn}, which is different from the
(Euclideanized) $X^0$. The identification of $\phi = X^0$
will be made eventually. 
From the expression
(\ref{centralcharge}) we conclude (c.f. (\ref{anom}))
that $\alpha_{0z} \sim \epsilon $ to leading order in
perturbation theory in $\epsilon$ and $u_z$, 
to which we restrict ourselves here.

We next remark that, as the analysis of~\cite{ellis96} indicates,
the $X^0$-dependent field operators
$\Theta_\epsilon (X^0)$ scale as follows with $\epsilon$:
$\Theta_\epsilon(X^0) \sim e^{-\epsilon X^0}
\Theta(X^0)$, where $\Theta(X^0)$ is a Heaviside step function
without any field content, evaluated in the limit $\epsilon \rightarrow 0^+$.
The bulk deformations, therefore, yield the following
$\sigma$-model terms: 
\begin{equation}
\frac{1}{4\pi \ell_s^2} \epsilon {u}_{z} X^0
e^{\epsilon(\phi_{(0)} - X^0_{(0)})}\Theta(X^0_{(0)})~\int _\Sigma
\partial_\alpha X^0
\partial^\alpha z~
\label{bulksigma}
\end{equation}
where the subscripts $(0)$ denote world-sheet zero modes.

Upon the interpretation of the Liouville zero mode $\phi_{(0)}$ as
a (second)
time-like coordinate, the deformations (\ref{bulksigma}) yield
space-time metric
deformations (of the generalized space time
with two times).
The induced metric components
can be
interpreted~\cite{ellis96}
as expressing the distortion of the space-time
surrounding the recoiling $D$-brane soliton.

For clarity,
we now drop the subscripts $(0)$ for the rest of this paper,
and we work in a region of space-time
on the $D3$ brane such that $\epsilon (\phi - X^0)$ is finite
in the limit $\epsilon \rightarrow 0^+$~\footnote{Note that 
the eventual
identification $\phi=X^0$ is thereby consistent
with this feature, as yielding a null contribution.}.
The resulting space-time distortion is therefore
described by the metric elements
\begin{eqnarray}
&~& G_{\phi\phi} = -1, \qquad G_{zz} =1, \qquad 
G_{00}=1\, ({\rm Euclideanized~}X^0),
\nonumber \\
&~& G_{\phi z} = \epsilon {u}_{z}X^0 \Theta (X^0)~,
\label{gemetric}
\end{eqnarray}
where the index $\phi $ denotes Liouville `time', not to be confused,
at this stage, 
with the original Euclideanized time $X^0$ (index $0$).

The presence of $\Theta(X^0)$ functions (impulse) and
the fact that we are working in the region $z >0$
indicate that
the induced space-time is piecewise continuous. The
important implications for non-thermal particle production
and decoherence for a spectator low-energy field theory
in such space-times were discussed in~\cite{kanti98,ellis96}.

We now make some important 
remarks about the metric (\ref{gemetric}).
Upon performing the transformation
$\phi \rightarrow \phi - \frac{1}{2}\epsilon {u}_{z} X^0 z $, 
which from a world-sheet point of view,
under the identification 
of the Liouville mode $\phi$ with a local world-sheet scale~\cite{emn}, 
may be considered 
as a renormalization-scheme choice,  
the
line element corresponding to (\ref{gemetric}) 
becomes, for $u_z \ll 1, \, \epsilon zu_z \ll 1$~\cite{leonta}:
\begin{eqnarray}
&~&ds^2 =-d\phi^2 + \left(1
-{b'}^2 ~z^2\right)~(dX^0)^2
+ \left(1 + {b'}^2 ~(X^0)^2\right)~dz^2 - 2{u}_{z}~z~dX^0~d\phi~,
\nonumber \\
&~& b'=\frac{1}{2}\epsilon {u}_{z}, \qquad u_z = g_s |\Delta p_z|/M_s
\label{bendinglineel3}
\end{eqnarray}
where $\Delta p_z$ 
is the momentum transfer along the bulk direction.

For our purposes of eventually applying the results presented here 
to models of space-time foam,
in which virtual $D$-brane configurations are emerging out of the vacuum, 
we are interested in the case in which 
there is no world-sheet
tree level momentum transfer. This naively corresponds to the case
of static intersecting branes. However, the whole philosophy of
recoil~\cite{kogan,szabo} implies that even in that case
there are quantum fluctuations induced by {\it summing up genera} on the
world-sheet. The latter implies the existence of a statistical
distribution of logarithmic deformation couplings of Gaussian type
about a mean field value ${\overline u}^{c}_{z}=0$. Physically, 
the couplings
${u}_{z}$ represent recoil velocities of the intersecting 
branes,
hence the situation of a Gaussian fluctuation about a zero mean
value represents the effects of quantum fluctuations about the
zero recoil velocity case, which may be considered as a quantum
correction to the static intersecting brane case. Such Gaussian
quantum fluctuations arise quite naturally by summing up higher
world-sheet topologies~\cite{szabo}, in particular 
pinched annuli, which have been argued to be 
the dominant configurations, and were shown to 
exponentiate. 
We therefore consider
a statistical average
$<< \dots >>$ of the line element (\ref{bendinglineel3}):
\begin{eqnarray}
&~&<<ds^2>> =-d\phi^2 + \left(1
-\frac{1}{4}\epsilon^2<<{u}^2_{z}>>~z^2\right)~(dX^0)^2 0
+ \nonumber \\
&~& \left(1 + \frac{1}{4}\epsilon ^2 <<{u}^2_{z}>>~(X^0)^2\right)~dz^2
- 2<<{u}_{z}>>~z~dX^0d\phi~, \nonumber \\
\label{bendinglineel2}
\end{eqnarray}
where
\begin{equation}
<< \cdots >>=\int _{-\infty}^{+\infty}d{u}_{z}
\left(\sqrt{\pi}\Gamma\right)^{-1}
(\cdots) e^{-{u}_{z}^2/\Gamma^2} \label{gauss}
\end{equation}
where the width $\Gamma$ has been calculated
in \cite{szabo}, after proper summation over world-sheet
genera, and is found to be
proportional to the string coupling $g_s$.

From (\ref{gauss})
the average line element
$ds^2$ becomes:
\begin{eqnarray}
&~&<<ds^2>> =-d\phi^2 + \left(1
-{b'}^2 ~z^2\right)~(dX^0)^2 
+ \left(1 + {b'}^2 ~(X^0)^2\right)~dz^2,
\nonumber \\
&~& b'=\frac{1}{2}\epsilon \Gamma
\label{bendinglineel3ab}
\end{eqnarray}
The definition of $b'$ comes from evaluating the quantity 
$<<{u}_{z}^2>>$ using the statistical distribution (\ref{gauss}).
Thus, in that case, averaging over quantum fluctuations leads to a
metric of the form (\ref{bendinglineel3}), but with a parameter
$b'$ much smaller, being determined by the width (uncertainty)
of the pertinent quantum fluctuations~\cite{szabo}.
The metric (\ref{bendinglineel3ab}) 
is exact, in contrast to the metric (\ref{bendinglineel3})
which was derived for $z << 1/b'$.

The width $b'$ expresses a momentum uncertainty of the 
fluctuating $D$-particle. 
Such uncertainties in general depend
on the (kinetic) energy content of the non-relativistic 
heavy $D$-particle defect. Even in isotropic situations, in 
which there is no tree-level average momentum for the $D$-particle,
${\overline u}^c=0$ such an energy dependence is non-trivial.  
For details we refer the reader to ref.~\cite{szabo}.  
For our purposes below we simply quote the result
for low energies $E$ compared to 
string scale $M_s$. In this 
case, the parameter $b'=b'(E)$ is given by:
\begin{eqnarray}
{b'}^2(E)=\epsilon^2 g_s^2\Bigl(1 - \frac{285}{18}g_s^2\frac{E}{M_D}\Bigr) 
\label{bdef1} 
\end{eqnarray}
where $E$ denotes the {\it kinetic} energy of the recoiling 
$D$-particle, 
$g_s$ is the string coupling, assumed weak, 
and $M_D=M_s/g_s$ is the $D$-particle mass, which is formally 
derived in the logarithmic conformal field theory approach 
from energy-momentum conservation~\cite{szabo}. 

It should be remarked at this point that,
throughout this work,  
we shall be working 
with $g_s $ small but {\it finite}. 
The limit 
$g_s \rightarrow 0$ will not be considered, given that when $g_s =0$ 
the mass of the $D$-particle $M_D \rightarrow \infty$, and thus
the recoil is absent, but on the other hand the curvature
of the surrounding spacetime, as a result of the immense mass
of the $D$-particle, should be taken into account. In that limit
the scale $b'(E) \rightarrow 0$, and thus one can no longer 
consider distances sufficiently far away from the center of the infinite 
gravitational attraction so that the Schwarzschild curvature
effects of the $D$ particle could be ignored.

As one observes from (\ref{bdef1}), the value of $b'$
decreases with increasing energy, and formally vanishes 
when the energy is close to $M_s$. 
We should note, however, that the above expression (\ref{bdef1})
pertains strictly to slowly moving strings, i.e. $E \ll M_s$.
In general, for arbitrary energies (including intermediate ones, which we shall
be interested in below), the precise 
expression for $b'(E)$ is not known at present.
For our purposes, 
we shall assume that $b'(E)$ decreases with increasing
$E$ for all energies. This will be 
justified later on.

An important feature of the line element (\ref{bendinglineel3ab}) is
the existence of an {\it horizon} at $z=1/b'$ for {\it Euclidean}
Neumann coordinates $X^0$. 
The presence of an horizon raises the issue of how one
could analytically continue so as to pass to the space beyond the horizon.
The simplest way, compatible, as we shall show later with the low-energy
Einstein's equations, is to take the absolute value of $1-{b'}^2z^2$
in the metric element (\ref{bendinglineel3ab}).
We therefore consider the following metric defined in all space $z \in R$:
\begin{eqnarray}
&~&ds^2 =-d\phi^2 + \left|1
-{b'}^2\, z^2\right|~(dX^0)^2
+ \left(1 + {b'}^2 ~(X^0)^2\right)~dz^2,
\label{bendinglineel3b}
\end{eqnarray}
For small $b'$, which is the case studied here,
and for Euclidean
Neumann coordinates $X^0$, the scale factor in front of the
$dz^2$ term does not introduce any singular behaviour, and hence
for all qualitative purposes we may study the following metric element:
\begin{eqnarray}
&~&ds^2 = -d\phi^2 + \left|1
-{b'}^2 ~z^2\right|~(dX^0)^2
+ ~dz^2,
\label{bendinglineel4}
\end{eqnarray}
which is expected to share all the qualitative
features of the full metric (\ref{bendinglineel3b})
induced by the recoil process in the case of an uncompactified
`bulk' Dirichlet dimension $z$ we restrict ourselves
here. 

The extension of (\ref{bendinglineel4}) to four dimensions
is straightforward, and has been made in~\cite{recoil}. 
In this case one views the $D$-particle as a zero-(spatial)-dimensional `wall' 
embedded in a three-dimensional `bulk' space.  
Here the terminology
`bulk' pertains to the longitudinal dimensions of the D3 brane 
in figure \ref{branefig}, and 
should {\it not} be confused with the real bulk transverse dimensions 
of the D3 brane domain wall. 
The four-dimensional analogue of (\ref{bendinglineel4}) 
is obtained upon making the following substitutions 
in the expression for the line element: 
\begin{eqnarray}
&~& z^2 \rightarrow r^2 = \sum_{i=1}^3 x_i^2  \nonumber \\
&~& dz^2 \rightarrow \sum_{i=1}^3 dx_i^2 
\end{eqnarray} 

An important comment concerns the relation of the Liouville
mode $\phi$ with the time $t$, under (\ref{time}). 
As already mentioned, 
in Liouville strings~\cite{ddk}, for which (\ref{anomcoeff}),
(\ref{rwliouv}) 
are valid, 
there is an implicit normalization of the Liouville mode $\phi$, which 
yields 
a canonical kinetic term in the $\sigma$-model action.
This normalization implies that 
the world-sheet zero mode of the Liouville field $\phi$ is connected 
to the area of the world-sheet $(L/a)^2$
via:
\begin{equation}
     \phi \sim Q_b\, {\rm log}(L/a)^2 \sim \epsilon^{-1} 
\label{areaphi}
\end{equation}
where $Q_b$ is the square root of the 
central charge deficit of the bulk world-sheet theory,
which in our case is of order $\epsilon$ (\ref{centralcharge}), 
given that the only source of non-criticality 
is the recoil of the $D$-particle. 

From (\ref{areaphi}) we thus observe that, 
for large times after the impulse, 
the Liouville mode and the target time 
scale similarly.  
This prompts one to identify these two fields. 
This is a crucial ingredient of the approach 
adopted in \cite{emn,recoil}, which is different from the treatment 
in \cite{leonta} where the Liouville mode had been fixed
to a certain value. 
Setting $\phi=X^0=t$ implies, on account of (\ref{edef}),(\ref{areaphi}), 
that 
$\epsilon^{-1}$ will scale as the physical time $t$~\cite{kogan,ellis96},
for large times after the impulse where our perturbative world-sheet  
approach is valid. This was expected from (\ref{time}), and thus 
it provides a nice consistency check of the approach. 

This identification, implies, that inside the initial horizon 
\begin{equation}\label{horizon} 
r < t/b'(E)\, , 
\end{equation}
where we have redefined 
$b'$ to be: 
\begin{eqnarray}
{b'}^2(E)=g_s^2\Bigl(1 - \frac{285}{18}g_s^2\frac{E}{M_D}\Bigr) 
\label{bdef} 
\end{eqnarray}
the induced space-time metric 
(\ref{bendinglineel4}), 
reads~\cite{recoil}:
\begin{eqnarray}
ds^2 = \frac{{b'}^2 r^2}{t^2}dt^2 - \sum_{i=1}^{3} dx_i^2~, \qquad r^2 =
\sum_{i=1}^{3} x_i^2 
\label{metricrecoil}
\end{eqnarray}
A careful analysis~\cite{recoil} shows that 
in this case the induced space time acquires a {\it bubble} structure,
which, as we shall discuss in the next section, matches smoothly 
a flat Minkowski space time outside the region bounded by 
the horizon (\ref{horizon}), (\ref{bdef}).  
An important point is that this space time is {\it unstable}.
This can be easily seen by examining the positive energy
conditions~\cite{recoil}, which are valid only in the 
interior of the bubble, being violated outside the region 
(\ref{horizon}).  For instance the weakest of these
conditions 
\begin{equation}
T_{\mu\nu}\zeta^\mu \zeta^\nu \ge 0, 
\end{equation}  
where $\zeta^\mu$ is an arbitrary  light-like (null) four vector, 
is easily seen to be valid only inside the region (\ref{horizon})
in the context of the space time (\ref{bendinglineel4}) 
under the constraint of 
the identification of time with the Liouville mode $\phi$.  
In the next section we shall try to remedie this
by demanding that the bubble space time is matched smoothly 
at the horizon boundary $r=t/b'(E)$. However, as we shall see
the instabilities will remain, in the sense that they will 
now appear as `tachyon-like' scalar excitations  of the string multiplet
inside the bubble.

It is the purpose of this work to 
discuss the physical implication of such instabilities.
As we have discussed above, and in \cite{leonta,recoil},  
the spacetime (\ref{metricrecoil}) 
may be considered as a {\it mean field} result of appropriate
resummation of quantum corrections for the collective coordinates
of the recoiling $D$-particle. To lowest order in a weak 
string ($g_s < 1$) $\sigma$-model perturbative framework, 
such quantum fluctuations may be obtained by resumming 
pinched annuli world-sheets, which can be shown 
to exponentiate~\cite{szabo}, thereby providing a 
Gaussian probability distribution 
over which one averages.
The metric (\ref{metricrecoil}) is the result of such an average,
and the eventual identification of the Liouville mode with the target 
time~\cite{emn}. The resulting instabilities, therefore, 
may not be considered as pathological, but a characteristic
feature of 
quantum fluctuating space times.

In order for the above procedure to be consistent, the resulting 
effective field theory must satisfy the appropriate $\sigma$-model 
conformal invariance conditions, which in a target-space framework 
correspond to appropriate equations of motion derived from a 
string-effective action. In this work we shall 
demonstrate that this is indeed the case, at least to lowest order 
in a Regge slope $\alpha '$ perturbative expansion where 
we restrict ourselves. 
We shall also study some physically important properties 
of the spacetime (\ref{metricrecoil}).
We shall argue that an inevitable result of 
the unstable nature of this spacetime
is the 
emission of high-energy radiation. 
Then we shall 
speculate on possible 
astrophysical applications of this phenomenon. 
Specifically, we shall argue that, as a consequence of the 
high-energy-photon emission from the unstable bubble, the neighbourhood of the 
recoiling $D$-particle defect may behave as a source of ultra-high-energy 
particles, which can reach the observation point, provided that the defect
lies at a distance from Earth which is within the respective mean-free paths
of the energetic particles. This effect, then, may 
have
a potential connection with the recently observed~\cite{gzkobs} apparent 
`violations' of Greisen-Zatsepin-Kuzmin (GZK) cutoff~\cite{gzkcutoff}.

\section{Dynamical Properties of the Recoil (Bubble) Spacetime}

We now commence our study of the most important properties
of the space time (\ref{metricrecoil}). 
It is our aim in this section to demonstrate 
that such space times are consistent solutions
of the ${\cal O}(\alpha ')$ conformal invariance conditions
of appropriately deformed closed-string sectors of the $\sigma$-model,
with target spacetimes dimensionally reduced to four dimensions.

To this end, we 
remark that, in a four-dimensional spacetime, obtained by appropriate compactification
of the higher-dimensional spacetime, where string theory lives, 
the string massless multiplet 
consists of 
a graviton field $g_{\mu\nu}$, a dilaton $\Phi$
and an antisymmetric tensor field $B_{\mu\nu}$, which in four dimensions
gives rise, through its field strength $H_{\mu\nu\rho}=3\partial_{[\mu}B_{\nu\rho]}$, to a pseudoscalar axion field $b$ (not to be confused
with the uncertainty parameter $b'$). The latter is defined as follows:
\begin{eqnarray}
H_{\mu\nu\rho}=\frac{1}{\sqrt{-g}}\epsilon_{\mu\nu\rho\sigma}\partial_\sigma b
\label{axion}
\end{eqnarray} 
What we shall argue below is that the spacetime (\ref{metricrecoil}) 
is compatible with the equations of motion obtained from a string  
effective action for the above fields. Equivalently, these 
equations are the $\sigma$-model conformal invariance conditions
to leading order in the Regge slope $\alpha'$.

We stress once again the fact that the spacetime metric 
(\ref{metricrecoil}) has been  derived upon the non-trivial assumption
that the target time is identified with the Liouville field~\cite{emn},
whose presence is necessitated by the recoil~\cite{recoil,kogan,szabo}.
The fact, to be demonstrated below, that this identification is consistent
with the $\sigma$-model conformal invariance conditions 
to ${\cal O}(\alpha')$,
is therefore a highly-non-trivial consistency check of the approach.

The components of the Ricci tensor for the metric (\ref{metricrecoil}) are:
\begin{eqnarray}
R_{00}=-\frac{2{b'}^2}{t^2}~, \qquad R_{ij}=\frac{\delta_{ij}}{r^2}-\frac{x_ix_j}{r^4} ~, \qquad i,j=1,2,3. 
\end{eqnarray}
The curvature scalar, on the other hand, reads:
\begin{eqnarray}
R=-\frac{4}{r^2} 
\end{eqnarray}
which is independent of $b'(E)$ and singular at 
the origin $r=0$ (initial position of the $D$-particle). 
Thus, we observe that the spacetime after the recoil acquires
a singularity. However, 
our analysis is only valid for distances $r$ larger than the 
Schwarzschild radius of the massive $D$-particle, and hence 
the locus of points $r=0$ cannot be studied at present
within the perturbative $\sigma$-model approach. It is therefore 
unclear whether the full stringy spacetime has a true singularity 
at $r=0$. 

The Einstein tensor $G^E_{\mu\nu} \equiv R_{\mu\nu}-\frac{1}{2}g_{\mu\nu}R$
has components:
\begin{eqnarray}
G^E_{00}=0~, \qquad G^E_{ij}=-\frac{\delta_{ij}}{r^2}-\frac{x_ix_j}{r^4}
\end{eqnarray}
The conformal invariance conditions for the graviton mode 
of the pertinent $\sigma$-model 
result in the following Einstein's equations as usual:
\begin{eqnarray}
G^E_{\mu\nu}=-{\cal T}_{\mu\nu} 
\end{eqnarray}
where ${\cal T}_{\mu\nu}$ contains contributions from string matter, 
which in our case includes 
dilaton and 
antisymmetric tensor (axion) fields, and probably cosmological constant
terms (which will turn out to be zero in our case, as we shall see later on).
As we shall also show, there are tachyonic modes necessarily present,
which, however, are not the ordinary flat-spacetime Bosonic 
string tachyons. In fact,
despite the fact that so far we have dealt explicitly with bosonic actions, 
our approach is straightforwardly extendible 
to the bosonic part of  superstring
effective actions. In that case, ordinary tachyons are absent from the
string spectrum. However, our type of tachyonic modes, will still be present
in that case, because as we shall argue later, such modes
simply indicate an instability of the spacetime (\ref{metricrecoil}). 

We find it convenient to use a redefined stress energy tensor 
${\cal T}'_{\mu\nu} \equiv {\cal T}_{\mu\nu}-\frac{1}{2}g_{\mu\nu}{\cal T}_\alpha^\alpha$,
in terms of which Eisntein's equations become:
\begin{eqnarray}\label{einst} 
R_{\mu\nu}=-{\cal T}'_{\mu\nu} 
\end{eqnarray}
We also assume that on the D3 brane world
the dilaton field is constant. This is a desirable
feature for late times, where our approach is valid.
In certain cosmological models this assumption may be relaxed, 
in which case a time-dependent dilaton field
plays the r\^ole of a quintessence field, 
as we shall discuss in section 10.

The stress tensor ${\cal T}'_{\mu\nu}$ for the case of tachyon and 
axion fields reads: 
\begin{eqnarray}
{\cal T}'_{\mu\nu}= \partial_\mu T \partial_\nu T 
+ \partial_\mu b \partial_\nu b - g_{\mu\nu}V(T) 
\label{stress}
\end{eqnarray}
where $V(T)$ is a potential for the tachyonic mode $T$. 
The fact that the axion field $b$ 
does not have a potential is dictated by the abelian gauge symmetry 
of string effective actions, according to which they depend only on 
the antisymmetric-field strength $H_{\mu\nu\rho}$ and {\it not} on the 
field $B_{\mu\nu}$. 
Below we shall show that indeed the field $T$
acquires a tachyonic mass, which however, in contrast to the flat-space
time Bosonic string theory tachyons, depends on the parameter $b'(E)$.

In addition to (\ref{einst}), one has the conformal invariance 
conditions for the tachyon and axion fields, to ${\cal O}(\alpha ')$:
\begin{eqnarray} 
\partial^2 T = -V'(T)~, \qquad \partial^2 b = 0 
\label{axiontachyon}
\end{eqnarray}
where the prime denotes differentiation with respect to $T(x_i,t)$.  

\begin{figure}
\epsfxsize=3.5in
\bigskip
\centerline{\epsffile{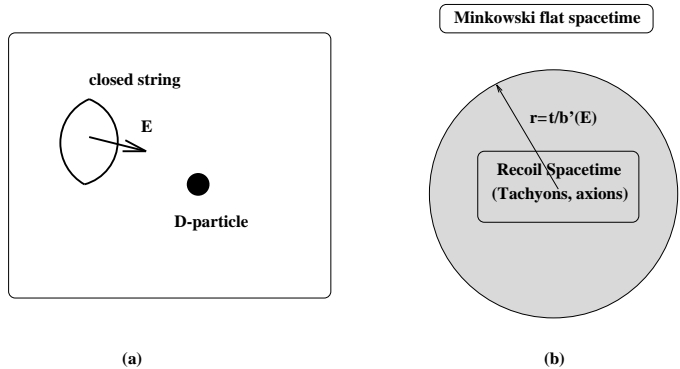}}
\vspace{0.2in}
\caption{(a) Scattering of a closed string mode 
(of energy E and non-trivial momentum) off a $D$-particle 
embedded in a four-dimensional spacetime (obtained by compactification 
of a higher-dimensional stringy spacetime). (b) after the recoil: 
formation of effective bubble, 
in the interior of which
the tachyonic modes 
and axion fields are confined. The  exterior geometry is the 
flat Minkowski.
\label{fig:bubble}}
\end{figure}

A solution to (\ref{einst}), (\ref{axiontachyon}) is given by:
\begin{eqnarray}
T(x_i,t)= {\rm ln}r~, \qquad b(x_i,t)=b'(E){\rm ln}t 
\label{solution}
\end{eqnarray}
provided that the tachyon potential $V(T)$ is:
\begin{eqnarray}
V(T) = -{\rm exp}\Bigl(-2T(r)\Bigr)
\end{eqnarray}
Naively, if the solution is extended to all space, we observe that 
the matter diverges logarithmically. To remedie this fact
we restrict the above 
solution to the range $r \le t/b'(E)$, and thus we enclose it in a 
{\it bubble} of time-dependent radius $t/b'(E)$. 
Outside the bubble we demand the spacetime to be the flat Minkowski
spacetime, and thus the above upper limit in $r$, $t/b'(E)$  
is the locus of points
at which the temporal component of the metric (\ref{metricrecoil}) 
becomes unity, and this allows an appropriate matching of the interior
and exterior geometries. 
As we shall see later on, 
a non-trivial consistency check of this matching 
will be provided {\it dynamically} 
by an explicit study of the scattering of 
test particles off the bubble. 
In this way the phenomenologically 
unwanted tachyon and probably axion fields (obtained
from the antisymmetric tensor field of the string) are 
confined inside the bubble of radius $t/b'(E)$ (cf. figure \ref{fig:bubble}).

In arriving at the above solution we have restricted ourselves to 
${\cal O}(\alpha')$ because we have ignored terms of higher order
in curvature $R$. To justify such an approximation it suffices to 
note that the ratio of the leading terms to the next to leading ones 
is:
\begin{eqnarray}\label{ho} 
(R/M_s)^2/R = {\cal O}\Bigl({b'}^2\Bigr)={\cal O}\Bigl(g_s^2(1-
g_s^2\frac{285}{18}\frac{E}{M_D})\Bigr)
\end{eqnarray}
provided one does not 
approach the singularity at $r=0$, which is in fact consistent 
with the regime of validity of the logarithmic conformal field 
theory~\cite{kogan,szabo}, i.e. 
distances far away from the defect, 
and times long enough after its collision with the string. 
This implies that our analysis is restricted near the 
boundary of the bubble, which will be 
sufficient for our purposes in this work. 
Since the string coupling is assumed weak, it 
is evident from 
(\ref{ho}) that the approximation of neglecting the higher-curvature
terms is satisfactory near the boundary of the bubble. Then, 
from the decreasing behaviour of $b'(E)$ with increasing energy, which, as 
mentioned previously, is assumed here even for intermediate energies, 
it follows that this approximation becomes 
even better for higher energy scales appropriate for the early stages 
of the universe.

A second important remark concerns the fact that in
the analysis leading to the metric (\ref{metricrecoil}) 
we have treated the spacetime surrounding the $D$-particle defect
as initially (i.e. before the collision with the string) flat.
However, even an initially at rest $D$-particle, 
being a very massive one of mass $M_D=M_s/g_s$ 
would naturally curve the spacetime around it, 
producing a Schwarzschild radius
$r_S=\ell_P^2/g_s\ell_s$, where $\ell_P$ denotes the 
four-dimensional Planck length 
and $\ell_s=1/M_s$ the string length. 
From our discussion above, the radius of the bubble of figure \ref{fig:bubble}
is $r_b=\ell_s/g_s $. For consistency of our approach, the approximation of 
treating the spacetime as initially flat implies that 
we work at distances considerably larger than the Schwarzschild radius,
so as the general relativistic effects due to the mass of the $D$-particle 
could be safely ignored. This implies that the radius of the bubble 
must be considerably larger than $r_S$, i.e. 
\begin{eqnarray}\label{condgs}
   1 \gg r_S/r_b = \Bigl(\ell_P/\ell_s\Bigr)^2  
\end{eqnarray}
{}From the modern view point of string/$D$-brane theory, the string length 
may not be necessarily of comparable order as the Planck length,
but actually could be 
larger~\footnote{For instance, in the brane-world models of 
\cite{RS}, where our world is the domain wall depicted in figure
\ref{branefig}, the bulk space time has a warped factor, characterised by a 
parameter $\kappa$, in such a way that the four-dimensional 
Planck Mass is of order $M_P^2 \sim M_s^3/\kappa$. 
If one assumes that closed strings are propagating
along the transverse directions 
of the brane of figure \ref{branefig}, 
then string effective action methods in the bulk, 
including ${\cal O}({\alpha '}^2)$ corrections,  
lead to~\cite{rizos} $\kappa \sim g_s M_s $. This implies that, 
in such models, the condition  
(\ref{condgs}) is satisfied upon 
the assumption of weak string couplings $g_s \ll 1$.}.
Thus, the above condition seems consistent. 

Notice also that the matching with the flat Minkowski spacetime in the 
exterior geometry is possible because the matter energy density 
${\cal T}_{00}$  and energy flow ${\cal T}_{0i}$ 
in the interior of the bubble of figure \ref{fig:bubble} are {\it both zero}:
\begin{eqnarray}
{\cal T}_{00}={\cal T}_{0i}= 0 
\end{eqnarray}
and thus there is no radiation coming out of or flowing into the bubble. 

We next compute the mass squared of the field $T$.
To this end, we shall consider the fluctuations $\delta T$ 
of the tachyonic mode 
$T$ around the classical solution $T_{\rm cl}$ 
(\ref{solution}) in the interior 
of the bubble, but close to its boundary, where the spacetime 
approaches the Minkowski flat geometry. From (\ref{axiontachyon}) 
we then have: 
\begin{eqnarray}
\partial^2 \delta T - 4e^{-2T_{\rm cl}}\delta T \Bigl|_{r\rightarrow t/b'(E)}=0 
\end{eqnarray}
from which we obtain a mass squared term of the form:
\begin{eqnarray}\label{mass} 
  m^2 =-4\frac{{b'}^2(E)}{t^2}~. 
\end{eqnarray} 
The negative value indicates, of course, the fact that the field $T$ is 
tachyonic, but the interesting issue here is that the induced mass
depends on the parameter $b'(E)$, and hence on the initial energy 
data of the incident string. 
 
A remark we would like to make at this point concerns the time-dependence
of the mass (\ref{mass}). As one observes from (\ref{mass}), the 
tachyon field will eventually disappear from the spectrum 
(as it becomes massless) asymptotically in time $t$. This fact comes from the 
specific form of the metric (\ref{metricrecoil}). 
From a non-critical string view point, discussed in section 2,
such time-dependent mass terms appearing inside 
the effective tachyon potential
$V(T)$, upon expanding in powers of the field $T$, 
will be considered as part of the central charge deficit $Q^2$,
which thus relaxes to zero asymptotically in time $t$ as 
\begin{equation}
     Q^2 \sim \frac{1}{t^2}
\label{tachccd}
\end{equation}
in agreement with (\ref{centralcharge}), 
(\ref{time})~\footnote{Due to the fact that these effects
occur in the interior of the bubble, whose radius is a 
few orders of magnitude larger than the Planck length, 
and taking into account the rare distribution of 
$D$-particles in the Universe, as imposed by
the phenomenology of the model, 
discussed in section 8, we remark that 
there will be no appreciable contributions to the vacuum energy 
of the Universe in a global scale
from the above phenomena. This could be a feature
only of certain cosmological models, to be discussed briefly in section 10.}. 
Notice that above we have obtained this result 
in an ``apparently conventional'' field-theory framework,
by matching our {\it unstable bubble}, 
bounded in space by a time 
dependent horizon (\ref{horizon}), with the flat Minkowski
space time in the exterior. In this way, 
the non-equilibrium nature of the underlying 
non-critical string theory, discussed in 
section 2, is hidden in this matching procedure 
involving an unstable spacetime. We think that this 
is an interesting and non-trivial 
feature of the model. 

However, as argued in \cite{recoil} quantum effects will eventually
stop the expansion of the bubble and may even force it to contract.
In this sense, the value of the mass of the tachyonic mode (\ref{mass}) 
will remain finite and negative, and will never relax to zero. 
Such effects can be understood in our context by the 
{\it breathing nature} of the evolution of the Liouville
field implied by the contour of integration depicted in 
fig. \ref{dollar}. The infrared world-sheet fixed point
(corresponding to an infinite area $A \to \infty$)  
acts as a `bounce'
point for this evolution~\cite{emn,kogan2}.

An equivalent way of seeing this is to observe 
that 
the explicit time-$t$ dependence in the temporal component of the metric 
(\ref{metricrecoil}) may be absorbed by a redefinition of the time: 
$dt'/l_s=dt/t$. In that case, the metric reads:
\begin{eqnarray}
   ds^2={b'}^2(E)r^2 d{t'}^2 - \sum_{i=1}^{3} dx_i^2
\end{eqnarray}
Under this redefinition, the bubble solution remain, but now the 
bubble appears to be independent of time, with its radius being 
$r=1/b'(E)$, and the mass squared of the tachyon being $m^2=-4{b'}^2$.
From that we observe that the tachyon mass remains 
$b'(E)$-dependent and finite. 
This argument supports the fact that even in the initial 
coordinate system (\ref{metricrecoil}) the time $t$ cannot be such so as 
to eliminate the $b'(E)$ dependence of the tachyon mass. 
This system of coordinates corresponds to 
a frame in which the bubble appears static, hence it corresponds 
in some sense to a \emph{comoving frame}. Therefore the mass
$m^2=-4{b'}^2(E)$ is the rest mass of the tachyonic mode.

Notice that the bounce (breathing bubble) picture is a feature
of the {\it microscopic} event horizons, being created
at a local scale around a recoiling $D$-particle.
In section 10 we shall consider also the recoil of the $D$-particle
in an expanding Friedman-Robertson-Walker Universe, in which case 
we shall study global effects of this recoil associated 
with the removal of possible cosmological horizons.
In such a context the relaxing-to-zero central charge deficit
(\ref{tachccd}),
which here has been associated with 
instabilities of the bubble space time, will be 
given physical significance as providing a global, time-dependent  
vacuum energy, relaxing-to-zero
a l\'a quintessence, for large times.

Some clarifying remarks are in order at this point concerning the 
nature of the tachyonic mode $T$. As we have just seen its mass is dependent
on the uncertainty scale $b'(E)$ of the $D$-particle, and hence is proportional
to the string coupling $g_s$ (cf. (\ref{bdef})). 
For this reason this tachyonic mode should be distinguished 
from the standard tachyon fields in flat-spacetime free Bosonic string theory. 
In fact, in our approach this tachyonic mode expresses simply
the {\it instability} of the {\it bubble} configuration, 
and will be present even in superstring effective field theories. 

As a result, from such considerations one may obtain an average
{\it lifetime} $\tau$ for the bubble:
\begin{eqnarray}\label{lifetime} 
   \tau = \frac{1}{2b'(E)}
\end{eqnarray}
in the ``comoving frame'', 
or 
\begin{eqnarray}\label{lifetime2} 
   \tau \sim l_s~ e^{1/2b'(E)}
\end{eqnarray} 
in the Liouville time frame,
where $\ell_s=M_s^{-1}$ is the string length scale. 
We remind the reader that 
for heavy $D$-particle 
defects (low kinetic recoil energy), we are dealing with here, 
$b' \sim g_s$ (\ref{bdef}). 
Working with weakly coupled strings, with couplings $g_s \ll 1$,  
we thus 
observe that the associated life times (\ref{lifetime2}) 
in the physical Liouville frame,
$\tau \sim \ell_s e^{1/2g_s}$,
are sufficiently larger than 
the minimum uncertainty times $\ell_s$, thereby justifying our
semiclassical perturbative treatment. We shall give some 
estimates in section 8, when we discuss possible physical
applications of the model.  

Some comments are now in order concerning 
the analysis in the Liouville-time frame. In our approach we 
prefer to work  in this frame, 
with the initial form of the metric 
(\ref{metricrecoil}), implied directly by the logarithmic
conformal field theory approach to recoil~\cite{kogan,recoil}.
This 
defines a {\it natural frame} for the definition of the {\it observable time}.
As we shall see this is also important when we discuss cosmological
implications in section 10. 
In this approach, the presence of the recoil degrees of fredom 
after a time, say, $t=0$ imply a breaking of the general coordinate 
invariance by the background, and also an irreversibility of 
time~\cite{recoil}. This comes from the fact that in our problem, time is 
a world-sheet renormalization group parameter (Liouville mode)~\cite{emn}, 
which 
is assumed irreversible, flowing towards a non-trivial infrared fixed
point 
of the world-sheet renormalization group 
of the Liouville 
$\sigma$-model~\cite{zam,kutasov,recoil}. As we have seen, 
in this frame the mass of the tachyon
appears time dependent, because, as we shall discuss below, the frame 
is not an inertial one, given that the spacetime of the bubble is a Rindler
\emph{accelerated spacetime}. 
There is one disadvantage, though, with the explicit 
use of the Liouville time $t$, 
in that it would require precise knowledge of the time moment at which 
the bubble was formed, and started decaying. Obviously, within 
a quantum string formalism such a process would occur within 
a (uncertainty) time  $\sim l_s$. Such issues  do not arise if one uses the 
``comoving'' time $t'$ instead.

\begin{figure}
\epsfxsize=2.5in
\bigskip
\centerline{\epsffile{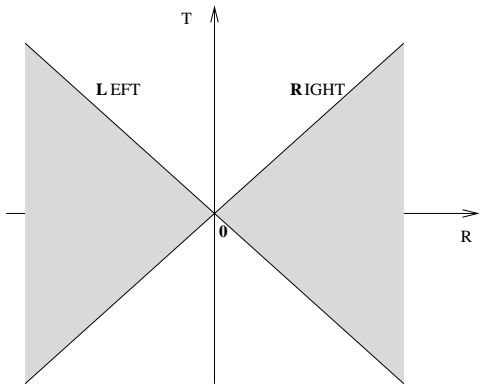}}
\vspace{0.2in}
\caption{Rindler wedge spacetime arising from the 
recoil of a $D$-particle, embedded in a 
four-dimensional spacetime, due to its scattering
with a closed string. The set of wedges (LEFT and RIGHT)
describe the spacetime for $t>0$. 
\label{fig:rindler}}
\end{figure}

\section{Thermal Effects and Emission of Radiation from the \\
Bubble Spacetime}

In this section we study issues associated with thermal radiation 
from our unstable bubble spacetime. 
To this end, we consider 
the metric (\ref{metricrecoil}), pass onto spherical polar coordinates $(r,\theta,\phi)$, and fix 
the angular part for convenience, as this does not modify 
the conclusions. We then 
perform the following coordinate transformations (from now on we work 
in units $\ell_s=1$):
\begin{eqnarray}\label{rindlertrnsf} 
R=r{\rm cosh}(b'(E){\rm ln}t)~, \qquad T=r{\rm sinh}(b'(E){\rm ln}t),
\end{eqnarray}
The transformation maps our space time to the right Rindler wedge (R) depicted
in figure \ref{fig:rindler}. The left wedge (L) is described 
by similar transformations up to a minus sign. 
In the $(R,T)$ coordinates the line element becomes:
\begin{eqnarray}
ds^2=dT^2-dR^2 - (R^2-T^2)d\Omega ^2 
\end{eqnarray}
where $d\Omega$ is the conventional solid angle.

From the above spacetime, we observe that for distances $R \gg T$ 
one recovers the flat Minkowski spacetime, whilst for distances
$R \sim T$ one obtains the {\it bubble} spacetime. 
Notice that $R^2-T^2=r^2$, and the interior of the bubble is defined 
by $r \le 1/b'(E)$ in comoving coordinates. 

An observer comoving with the expanding bubble, placed at position $r$,
is accelerated with respect to the $(R,T)$ frame, with proper 
acceleration $1/r$. According, then, to the standard analysis 
of accelerated observers~\cite{davies}, such an observer sees
the Minkowski vacuum (in the $(R,T$) coordinates)
as having a non-trivial {\it temperature} $T_{\rm bubble}$  
\begin{eqnarray} 
    T_{\rm bubble}=\frac{1}{2\pi r}, \qquad  0 < r < 1/b'(E)
\end{eqnarray}
The temperature for the inertial Rindler observer $T_{0}$
is:
\begin{eqnarray}\label{trindler}
T_0 =\sqrt{g_{00}}T_{\rm bubble}=\frac{b'(E)}{2\pi}
\end{eqnarray}
The presence of temperature is expected to imply a non-trivial proper 
entropy density $s$. For a massless scalar field, in our case the axion $b$, 
the entropy $S=\int d^3x \sqrt{-g} s$ 
in four spacetime dimensions is given by: 
\begin{eqnarray}
  S=\int d^3x \sqrt{-g} \frac{4}{3}\frac{\pi^2}{30}T_{0}^3 
\end{eqnarray}
From the bubble spacetime (\ref{metricrecoil}) one then obtains:
\begin{eqnarray}\label{entropy} 
S=\int d^3x \sqrt{-g} s=4\pi b'(E) \int _0^{1/b'(E)}dr~r^3  
\frac{4}{3}\frac{\pi^2}{30}\frac{{b'}^3(E)}{8\pi^3}  =\frac{1}{180} 
\end{eqnarray}
Thus the bubble carries non-trivial entropy, which turns out to 
be independent 
of $b'(E)$. The reader should not be alarmed by the apparent volume 
independence of the entropy, which at first sight 
would seem to contradict the fact 
that the entropy is an extensive quantity. In fact, there is no 
contradiction in our case, since 
there is only one scale in the problem, 
$b'(E)\ell_s$, and the volume of the bubble is itself expressed
in terms of this scale.

The presence of entropy production after the recoil implies 
{\it loss of information} which can be understood as follows: 
one starts from a pure state of a string striking a $D$ particle. 
There is no entropy in the initial configuration. 
After the moment of impact, the $D$-particle recoils, and because it is a 
heavy object it distorts the spacetime around it, 
producing the bubble phenomenon via 
its recoil excitation degrees of freedom.
Due to the finite lifetime of the bubble, 
the entropy (\ref{entropy}) will be released to the external space,
implying information encoded in the recoil degrees of freedom, which 
are unmeasurable by an asymptotic observer. 
This picture is consistent with the loss of conformal invariance
of the underlying $\sigma$-model, and the irreversible 
world-sheet renormalization-group 
flow of the recoiling system, as discussed in \cite{recoil,emn}, 
upon the identification of the Liouville mode with the target time. 

It should be remarked at this point 
that the entropy (\ref{entropy}) 
pertains strictly to scalar fields that live {\it 
inside the bubble}. There is no crossing of the surface of the bubble 
by the interior fields in our construction, at least classically
(as we shall discuss below there is a quantum-mechanical 
escape probability). 
On the other hand, it should be 
noticed that for a field 
which lives in the {\it exterior} of the bubble, 
there appears to be 
loss of information in the sense 
that the exterior particle degrees of freedom 
may enter the bubble and be captured, as we shall discussed 
in the next section. 
Thus, 
an asymptotic
observer, far away from the bubble, will necessarily {\it trace out}
such (unobserved) 
degrees of freedom in a density matrix formalism, and in this 
sense the resulting entropy,
pertaining to such degrees of freedom,
will be proportional 
to the {\it area} of the bubble and {\it not its volume}~\cite{srednicki}.
We shall come back to this important point, in connection with 
emitted radiation from the bubble later on. 

\begin{figure}
\epsfxsize=2.5in
\bigskip
\centerline{\epsffile{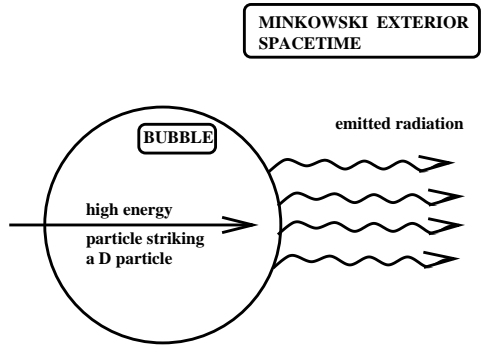}}
\vspace{0.2in}
\caption{Emitted radiation from the unstable 
bubble. The radiation is not isotropic, 
but most of it will be emitted in the forward direction,
parallel to that of the incident high-energy particle.
\label{fig:thermal}}
\end{figure}

The presence of temperature $T_0$ (\ref{trindler}) implies the 
emission of radiation from the bubble (see figure \ref{fig:thermal}), 
which can be 
read off from the Stefan-Boltzman law $\sigma T_0^4$, $\sigma = \pi^2/60$ 
(in units $\hbar=c=k_B=1$). Given that the area is $4\pi {b'}^{-2}(E)$,
and that the lifetime of the bubble is estimated from (\ref{lifetime}),
one observes that during the life time $\tau$, the following 
amount of energy is released in the form of radiation:
\begin{eqnarray}\label{energyrad} 
    E_{{\rm rad}} \sim 4\pi {b'}^{-2}(E) \tau \sigma T_0^4 =
\frac{1}{480\pi}b'(E)M_s=\frac{g_s}{480\pi}b'(E)M_D  
\end{eqnarray} 
It is interesting to observe that the same amount of energy 
represents the {\it thermal} energy of the axion field in the interior 
of the bubble:
\begin{eqnarray}
  E_{{\rm th/axion}}=\int d^3x \sqrt{-g}\frac{\pi^2}{30}T_0^4 
= \frac{g_s}{480\pi}b'(E)M_D
\end{eqnarray}
From (\ref{bdef}), and taking into account that in the effective 
field-theory 
limit we are working here, $E < M_s$, and that $g_s \ll 1$ 
in our weak string framework, 
one observes that the energy $E_{{\rm rad}}=E_{{\rm th/axion}} < M_D$.

From energy conservation then, which, notably, 
is shown to be valid rigorously 
in the context of our logarithmic conformal field theory 
(stringy) recoil framework~\cite{szabo},
one obtains:
\begin{eqnarray}
E_{\rm in} + M_D = M_D + E + E_{\rm th/axion} 
\end{eqnarray}
where $E_{\rm in}$ denotes the total energy of the {\it incident} 
particle/string. 
From this, one thus sees that there is a {\it threshold} 
for bubble formation:
\begin{eqnarray}\label{threshold} 
E^{\rm threshold} = E_{\rm rad}= E_{\rm th/axion}
\end{eqnarray}
From these considerations, 
one observes that the radiation energy {\it will not cause any 
mass loss} of the $D$-particle, since all the thermal axion energy
accounts for that. Hence, despite the instability 
of the bubble, the stability of the $D$-particle defect is not affected.
This will be important for our physical applications, to be discussed 
later on.

Notice that the energy release (\ref{energyrad}) 
cannot make up for the 
maximum of the thermal energy expected from Wien's law 
$\lambda T_{\rm max} = {\rm const}$, where $\lambda$ the wavelength
of radiation. Thus the resulting photon spectrum is {\it not thermal},
and hence one can only get an estimate for the energies of the 
emitted radiation. The alert reader might then object to our
previous use of the black body (or, in general, equilibrium) laws.
In fact their use is indicative, and they can only give
qualitative results (e.g. in our case we only obtain a  
tail of the thermal distribution).

To recapitulate, the physics behind the above properties
can be summarized as follows: one needs a highly-energetic 
incident particle of energy $E_{\rm in}> E^{\rm threshold}$, which strikes
a $D$-particle, and forms a bubble; the bubble radiates an amount 
of energy 
$E^{\rm rad}$ (\ref{energyrad}) distributed appropriately among 
the various photons. The emitted radiation will not be isotropic 
as a result of (spatial) momentum conservation
(see figure \ref{fig:thermal}). This will be useful 
in physical applications. 
The fact that a particle, entering 
and being captured by the bubble, will
cause the 
emission of radiation from the bubble
is nicely related to the existence of non-zero entropy 
measured by an asymptotic observer. 
This phenomenon is thus not dissimilar to the Hawking process of an
evaporating black hole, although in our case 
the bubble is {\it not} a black hole, neither the $D$-particle evaporates,
as we discussed above. This picture is in agreement 
with the 
non-equilibrium Liouville string framework~\cite{recoil}, on which  
the approach is based.

\section{Motion of Particles in the Bubble Spacetime}

We shall now analyze the motion of a particle 
in the bubble spacetime (\ref{metricrecoil}). 
For convenience we shall work at the equator of the three sphere 
(fixed angle $\theta =\pi/2$). The Lagrangian of the particle 
in the background spacetime (\ref{metricrecoil}) in comoving coordinates 
$(t',r, \phi)$ is:
\begin{eqnarray}
{\cal L}=\frac{1}{2} \Bigl(\frac{ds}{d\lambda}\Bigr)^2 =
\frac{{b'}^2(E)r^2}{2}\Bigl(\frac{dt'}{d\lambda}\Bigr)^2 
- \frac{1}{2}\Bigl(\frac{dr}{d\lambda}\Bigr)^2 - r^2\frac{1}{2}
\Bigl(\frac{d\phi}{d\lambda}\Bigr)^2 
\end{eqnarray}
Expressing the Lagrangian in  terms of the conserved angular momentum 
$L=r^2 (d\phi/d\lambda)^2$
and energy $E_{\rm in}={b'}^2(E)r^2 \tfrac{dt'}{d\lambda}$ we obtain:
\begin{eqnarray}
\frac{E_{\rm in}^2}{2{b'}^2(E)r^2}-\frac{1}{2}\Bigl(\frac{dr}{d\lambda}\Bigr)^2 
- \frac{L^2}{2r^2} =\frac{\mu^2}{2}
\end{eqnarray}
where 
$\mu=0$ for massless particles (e.g. photons), and $1$ for massive particles
(in which case the quantities $L$ and $E_{\rm in}$ 
are the corresponding quantities per 
unit mass).

Writing the equation of motion as:
\begin{eqnarray}\label{effpot}
  \frac{1}{2}\Bigl(\frac{dr}{d\lambda}\Bigr)^2+\frac{\mu^2}{2}  =
\frac{E_{\rm in}^2/{b'}^2(E) - L^2}{2r^2}
\end{eqnarray}
we can see that the impact parameter 
$L/E_{\rm in}$ must be smaller than $1/b'(E)$, for the equation to make sense.
This means that if $L/E_{\rm in} > 1/b'(E)$, then the particle 
will {\it necessarily} travel {\it outside} the bubble spacetime, 
which is thus a dynamical consistency check of our matching 
assumptions that the spacetime in the region $r \ge 1/b'(E)$ 
is the flat Minkowskian spacetime. Such outside particles 
will not be affected by the presence of the bubble, and their 
trajectory will be undisturbed, that predicted by special 
relativistic dynamics.

Below we shall study now the case of impact parameters 
$L/E_{\rm in} < 1/b'(E)$,
for massless and massive particles. In such a case, from (\ref{effpot}) 
the massless particle equation of motion reads: 
\begin{eqnarray} \label{plusminus} 
   r=r_0
{\rm exp}\Bigl(\pm\phi \sqrt{\frac{E_{\rm in}^2}{{b'}^2(E)L^2}-1}\Bigr)
\end{eqnarray}
where in the case of an incoming photon (from outside the bubble) 
$r_0=\frac{1}{b'(E)}$, and we have taken only the 
$(-)$ sign, because this is the only consistent choice. 
This shows that the massless particle will be
{\it captured} inside the bubble. 

From (\ref{effpot}) one observes that, 
in the case of a massive particle, there exist values of energy and 
angular momentum such that the particle can escape the bubble spacetime. 
This happens if 
the 
radial velocity on the boundary (as the particle attempts to escape)
is non-zero, which implies (we have re-instated the units of $M_s$ 
for clarity):
\begin{eqnarray}\label{escapecond} 
E_{\rm in}^2 - L^2{b'}^2(E)M_s^2 > M_s^2
\label{condition} 
\end{eqnarray}
This demonstrates that only highly energetic particles with energies
much higher than $M_s$ can escape the bubble spacetime.

\begin{figure}
\epsfxsize=2.5in
\bigskip
\centerline{\epsffile{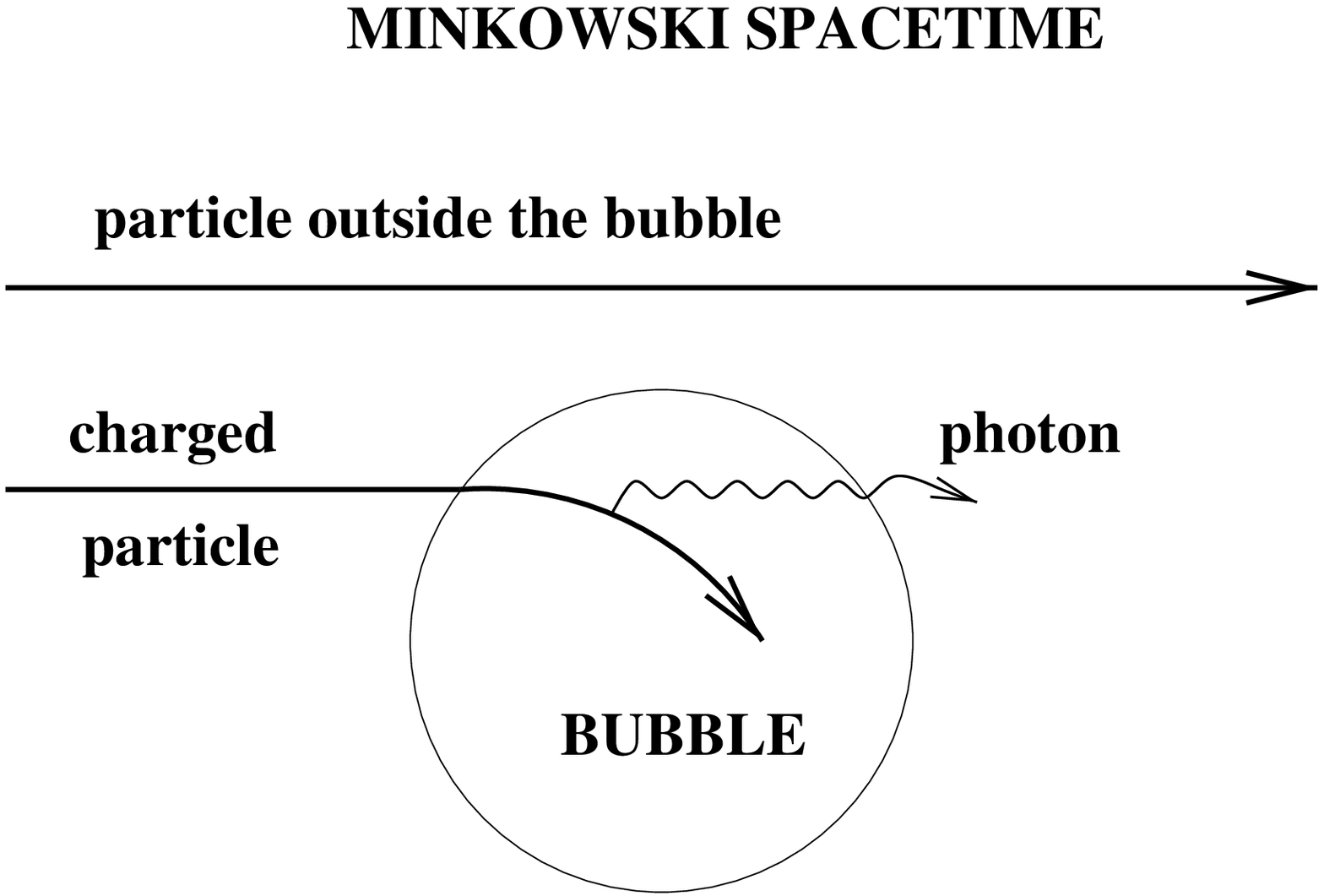}}
\vspace{0.2in}
\caption{Emission of photons due to non-uniform motion 
of a charged particle inside the bubble. The radiation 
can escape the bubble.
\label{fig:transition}}
\end{figure}

Once such a particle is electrically charged, its non-uniform 
(spiral) motion inside the bubble will  cause the emission of 
{\it radiation}. The latter will continue to carry angular momentum 
and energy 
of roughly the same order as that of the particle. Because 
the emission is now taken place at $r_0 < 1/{b'}(E)$ 
the positive sign in the exponent of equation (\ref{plusminus}) 
is also allowed, implying an escape possibility for the emitted 
photons (cf. figure \ref{fig:transition}). 
In addition, as we shall show in the next section, the 
bubble has a non-trivial (thermal) refractive index, and thus 
behaves as a medium. If there is a beam of charged particles 
entering the bubble within its short life time, then 
these particles will experience the (suppressed) phenomenon 
of {\it transition radiation}~\cite{trt}, i.e. the emission 
of photons accompanying 
an electrically-charged particle when  
crossing the interface separating two media with different
refractive indices (in our case, the interior of the bubble and the 
exterior Minkowski spacetime). 
A fraction of this radiation  
will also escape the bubble. 
Such phenomena, if true, will imply excess of photons
accompanying the charged particle. We shall present a more 
detailed discussion of these issues, and their 
potential experimental consequences, 
in a forthcoming publication.

It should be noticed that, if the condition (\ref{condition}) is satisfied, 
then the particle is {\it deflected} by an angle $\Delta\phi$ 
which can be computed in a standard way to be:
\begin{eqnarray}\label{deflection}  
\Delta\phi = \pi - \frac{2}{\sqrt{\frac{1}{\rho^2{b'}^2(E)}-1}}
{\rm arccosh}\Bigl(E_{\rm in}\sqrt{1-\rho^2{b'}^2(E)}\Bigr)~, \qquad \rho=L/E_{\rm in}={\rm impact~parameter}
\end{eqnarray}
{}From this we observe that, for fixed impact parameter $\rho$, 
highly energetic particles will not be deflected much, as 
should be expected.

As a final comment we mention that the scattering 
cross section $\sigma (E)$ 
for energies and/or angular momenta 
that violate the 
condition (\ref{condition}), which are of physical interest, 
is given by:
\begin{eqnarray}\label{crossection} 
\sigma (E) = \pi {b'}^{-2}(E) 
\end{eqnarray}
From the results of ref. \cite{szabo}, which are valid strictly 
only for low energies,
we observe that $b'(E)$ decreases with increasing energy $E$ (\ref{bdef}). 
One would expect intuitively that strings with higher energy 
would cause larger distortion of the spacetime surrounding the 
recoiling $D$-particle defect. This point of view is 
supported by  the above results if one 
extends the behaviour of 
$b'(E)$ encoded in (\ref{bdef}) to intermediate 
energies as well, and in fact to all energies (up to Planckian), 
because in that case the distortions of spacetime 
caused by strings with higher energy will correspond to 
formation of bubbles with bigger radii $1/b'(E)$. 

From (\ref{crossection}) we observe that 
higher energy strings would correspond to larger cross sections.
The important point to notice is that the cross section 
is non zero even for zero energy. This stems from (\ref{bdef}),
and is associated with the fact that $b'(E)$ is 
essentially a  {\it quantum uncertainty} 
in the momentum of the $D$-particle~\cite{szabo}, which is 
not zero even for vanishing incident energy $E$.  
Hence, the spatial uncertainty of the position of the 
$D$-particle, which is associated with the cross section, is not zero, 
but is bounded from below by the Heisenberg uncertainty principle,
which explains naturally the non-zero minimum value of $\sigma (E=0)$. 
It also explains the increasing behaviour of the 
cross section with increasing energy, given that the higher 
the energy is, the larger the uncertainty is expected to be. 

\section{Quantum Electrodynamics Effects inside the Bubble and 
Refractive Index for Photons}

So far our considerations have been classical. 
It is in this sense that we demonstrated capture of 
photons (and, in general, massless particles) 
in the interior of the bubble. 
The presence of 
finite temperature (\ref{trindler}) will create a 
non-trivial (thermal) vacuum, with broken Lorentz symmetry 
inside the bubble. In that case, it is known~\cite{pascual}
that the effective velocity of light, defined by the 
quantum propagator of photons, is modified in accordance 
with the fact that the finite temperature effects 
provide the notion of a medium. 

Specifically, the dispersion relation for a 
particle of mass $m$ in the non-trivial 
vacuum at temperature $T$ is: $E^2(p)=p^2 + f(p,T,m)$,
where $E$ is the energy and $p$ the momentum, and 
the function $f$ encodes the quantum effects of the 
vacuum polarization. The group velocity of the particle 
is then given by $v=\partial E(p)/\partial p$, and in general 
depends on $p$. 

In the case of photons in a 
non-trivial quantum electrodynamical vacuum, the function 
$f$ can be computed, to one loop, from the vacuum-polarization graph 
of the photon~\cite{pascual}. The latter
is of order $f \sim T^2e^2$, where $e$ is 
the electron charge~\footnote{Note that one can extend these results to 
incorporate all known particles in the standard model~\cite{pascual}.}. 

In our case, the induced temperature (\ref{trindler}) is much larger than the 
incident momentum $p$ of the photons. From the results of \cite{pascual}
in this case, the effective velocity of the photons 
inside the bubble is (in units of speed of light {\it in vacuo} $c=1$):
\begin{eqnarray}\label{index} 
v \sim \frac{p}{eT_0} \sim \frac{p}{e {b'}(E)}   
\end{eqnarray}
where $p$ is related to the energy $E$ of the photon 
via the (modified) dispersion relation. The photon in this case
becomes {\it effectively} massive, with mass $\mu \sim e {b'}(E) \ne 0$. 
It is interesting to note that, 
due to the extreme temperature effects, $p \ll T_0$, the resulting photon 
is considerably slowed down inside the bubble 
to non-relativistic velocities.

From our earlier discussion on massive particle trajectories
inside the bubble, and the 
fact that the quantum effects result in an effective photon mass,
one is tempted  to consider the
possibility of a photon escaping the bubble. 
However, 
from the condition (\ref{escapecond}) 
and the induced photon mass $\mu \sim e {b'}(E)$, 
it becomes evident 
that 
there is {\it no such possibility}.

Nevertheless, there is a non-zero 
probability of {\it quantum tunnelling} 
through the potential barrier.  
In this sense, part of the classically captured (incident) photons can 
escape.
Combined with the thermal slowing down (cf. (\ref{index})) of all photons 
inside the bubble, then, 
this will result in the appearance of 
{\it delays} in the respective arrival times of photon beams
from distant astrophysical sources in areas where there are 
$D$-particles. Because these delays will be associated with 
only part of the photon beam, the final effect will appear 
as a fluctuation in the arrival time.

The associated delay for a single photon, which   
passes through a region in space 
where there is one $D$-particle, and is assumed to escape 
through tunnelling, 
is estimated to be: 
\begin{eqnarray} \label{delay} 
   \Delta t \sim \frac{1}{b'(E)v(E)} = \frac{e}{p} 
\end{eqnarray}
where $\frac{1}{b'(E)}$ is the radius of the bubble, $e$ is the 
electron charge,
and this formula is applicable for $p \ll M_s$. For velocities
$p \sim M_s$ there are modifications, which however 
are not of interest
to us here. 

The existence of delay effects that depend on the energies of the photons
bares some resemblance to quantum spacetime 
effects, associated with induced refractive index for photons, discussed in 
\cite{sarkar}. As in those works, so in the present model there 
appears to be a non-trivial refractive 
index $n(E)=1/v(E)$ inside the bubble. However, 
there is an important difference, in that
here this is due to conventional quantum electrodynamical thermal effects. 
Because of this reason, the induced refractive index (\ref{index}) 
is reduced with increasing photon energies, in contrast to the 
effects of refs. \cite{sarkar}, where the quantum spacetime induced 
refractive index appears to increase with energy. 
However, the reader 
should bear in mind that in our model the existence of temperature is due to 
quantum stringy effects~\cite{recoil,szabo} associated with the 
recoil of the $D$-particle. In this sense there is, 
in our approach, a notion of quantum gravity, in similar spirit 
to the work of \cite{sarkar}.  

Unfortunately, for a very dilute gas of $D$-particles the maximal 
delay 
(\ref{delay}), corresponding to the less energetic 
observable photons (e.g. infrared background of 
energies 0.025 eV) is very small, or order $10^{-11}$ s.
Nevertheless, we should notice that if one considers
photon sources from distant astrophysical objects that 
are at distances corresponding to cosmological redshifts $z > 1$,
then in those areas of the relatively early universe, the density 
of $D$-particles might have been higher. 
One then can get an upper bound of such densities 
by considering the effect (\ref{delay})
as lying inside the present experimental 
errors.

\section{Possible Astrophysical Constraints on the Model: \\ Clusters 
of $D$-Particles in the Universe and GZK Cutoff} 

In this section we would like to make some speculative 
remarks on 
possible physical 
applications of the above phenomena. 
It is by no means {\it a priori} obvious that features of Liouville strings 
can be constrained (or even tested!) 
experimentally in the immediate future. 
However, the specific non-critical string/$D$-brane model under consideration
has the feature that 
it leads to emission 
of very high-energy radiation from the recoiling $D$-particles.
As we shall argue below this feature imposes severe restrictions
on the parameters of the model. 

Let us assume that there is a rare distribution of $D$-particles
in the inflated universe (although their density 
was much higher in the early universe). Due to this rare distribution, 
the concept of isotropy 
is not applicable. It is therefore possible that an isolated $D$-particle
lies between Earth and a distant galaxy.
As the galaxy emits particles, some highly energetic 
and weakly interacting ones, such as neutrinos, strike the 
$D$-particle and induce the formation 
of a bubble. The latter then emits radiation in the way 
explained above. The emitted photons, in the direction of the incident 
particles, will be highly energetic, of typical energy $b'(E)$. 
In general, such photons will interact with the background photons 
of either the microwave background radiation, or the infrared background,
to yield, say, $e^+e^-$ pairs. Because of this, if the $D$-particle 
lies far away from Earth, outside the average mean free path 
of such photons, the latter will not arrive on Earth. 
However, one may imagine a situation in which the isolated $D$-particle
lies within the above mean free path distance, which in the case 
of photons interacting with the infrared background 
is estimated to be of order of a few Mpc~\cite{protheroe}. 
Then, the weakly interacting incident particles that trigger the phenomenon
of bubble formation, e.g. 
neutrinos, will  
arrive at the location of the $D$-particle(s) undisturbed. 
In that case, the 
emitted high-energy radiation will reach the observation point, and 
in this sense the recoiling $D$-particle bubble {\it constitutes a novel 
and relatively nearby source of highly energetic photons}. Such scenaria 
may have applicability to the recently observed 
highly energetic 30 TeV photons that seem to violate the so-called
GZK cut-off~\cite{protheroe,gzkobs}. 

In a similar way, one can also extend the above discussion to incorporate
charged particles. Consider, for instance, a beam of protons 
emitted by a galaxy lying at cosmological distances, 
whose energies
do not exceed the GZK cut-off, and hence they arrive 
undisturbed
until the point where a $D$-particle defect lies.
One of the protons will then strike the $D$-particle
and create a bubble spacetime. As discussed previously, the proton
will be captured inside the bubble, since its 
energy does not satisfy the escape condition (\ref{escapecond}), 
as being less than $M_s$~\footnote{Here we assume that 
$M_s$ exceeds the GZK cut-off $10^{19} {\rm eV} (= 10^{10}$ GeV). 
If this is not the case, and one has a lower $M_s$, then such 
massive particles may escape. At any rate, 
our discussion does not 
depend upon this fact.}. 
The bubble will then radiate very high energy photons,
which can interact with the remaining protons 
in the beam, that fly outside the bubble, to create, say, 
protons and pions {\it etc}. The protons (or, in general, 
the particles) that emerge from 
such interactions   
will then be very energetic, and it is conceivable that their energies
can be of the order of the observed~\cite{gzkobs} $3 \times 10^{20}$ eV.
In this way, the region around the recoiling $D$-particle defect
acts as a source of ultra-high-energy cosmic rays, and if one 
assumes, as before, that the defect lies within the mean free path 
of a proton (from Earth), this can easily contribute to 
observed apparent ``violations'' of the GZK cut-off~\cite{gzkobs}. 

If the above scenario survives, it may then 
imply that there is no actual violation of Lorentz symmetry 
that is responsible for the  phenomenon, as claimed by a number 
of authors~\cite{gzk}, since in our bubble spacetime 
there is no such violation (except the trivial one due to temperature
effects inside the bubble). To put it in simple terms, 
in our scenario, the source of the ultra-high-energy 
cosmic rays is in the neighbourhood of the $D$-particle defect, 
which may lie at a much closer distance from Earth 
than one naively thought. 
Of course, it goes without saying that one cannot exclude 
the possibility that 
a peculiar  
combination of phenomena, involving the model discussed here
in conjunction with both conventional
and unconventional (spacetime foam) physics~\cite{sarkar,gzk}, 
might 
actually
lie behind such extreme astrophysical phenomena.

We shall now make use of the above considerations to 
constrain the model by employing 
current observational data related to ultra-high-energy cosmic 
rays~\cite{gzkobs}.
Alternatively, from an overly optimistic viewpoint, 
such a `phenomenology' may be seen as providing  
a means of possible experimental 
detection of the $D$-particles, if clusters of them really exist in
our galactic neighbourhood!
We will show that, under a few
assumptions, radiation from a physically acceptable population of $D$-particles
in the universe can be observed. Then, an excess of photons somewhere
in the observed spectrum of high energy cosmic rays, can be interpreted as
coming from this effect and, through that, from the branes. 
Since high energy photons
have a mean free path smaller than 100 Mpc, 
we shall consider here $D$-particles
located inside a volume of radius of order of 10 Mpc, around Earth. 
However, as we shall show below, 
compatibility with standard astrophysics implies that 
only a small part, if at all, of the observed 
ultra high energy cosmic ray events can be attributed to 
this kind of exotic physics.

To this end, 
let us consider a $D$-particle striken by a current of
highly-energetic weakly-interacting particles,
which can travel undisturbed at great distances.
We assume that the energy of the particle
is higher than the threshold energy (\ref{threshold})
for the formation of the bubble. The highest 
energy neutrinos from gamma-ray bursters 
(GRB's)~\cite{neutrino} can provide 
the high energy flux we need, 
as in the bursters there is a relatively significant production of 
particles whose energies 
can possibly be of 
order $10^{-2}g_sM_s$, where 
$g_s$ is assumed smaller than unity, for the validity of the 
string perturbation expansion. 
Our interest in neutrinos arises from the fact that these particles 
can travel
undisturbed over very large distances, creating significant 
uniform and isotropic flux. According to contemporary models~\cite{neutrino}, 
these neutrinos 
come with appreciable flux only up to energies
$10^{18}$eV, so our first assumption will be that the threshold
energy must be bounded from above by such a value. This, in turn, implies 
that 
the string length scale falls within the experimental sensitivity, 
if it is at most (c.f. (\ref{threshold}))
\begin{eqnarray}\label{stringscale}
  M_s \sim \frac{10^{11}~GeV}{g_s}
\end{eqnarray} 
Then, a few photons will be emitted, with an energy spectrum
in a narrow window below $E_{rad}$, whose highest value has been assumed to be
about $10^{18}$eV. Note that, if one uses 
$g_s\sim 10^{-2}$, which is 
small, as needed for
the validity of our string perturbation expansion, but not unjustifiably 
small,
we observe   
that the smallest detectable string length, in this approach,
is $l_s \sim 10^{-27} {\rm cm}$, in which case 
the bubble's lifetime 
(\ref{lifetime2}) is $\sim 10^{-16}$ sec. This is much larger 
than the minimum string time scale $\ell_s/c \sim 10^{-38}$ sec,
which implies that our perturbative approach is consistent in this case.

We now turn to an estimation of the distribution 
of $D$-particles in the universe which can
produce sufficient rates of photons on Earth.
We assume a {\it uniform} and {\it isotropic}
flux $J_{\rm he}$ for the highly-energetic particles,
and a concentration of $D$-particles $n_D$ in a region of
linear dimension $\ell$. Then, it is evident that the
flux of the incident particles, which will strike the
$D$-particles, is $J_{\rm he}(1-e^{-\ell \sigma n_D})$
where $\sigma$ is the effective cross section of the
particle/$D$-brane scattering :
\begin{eqnarray}\label{cross}
  \sigma \sim l_s^{2} = (\frac{10^{11}}{g_s}GeV)^{-2} = g_s^{2}10^{-50} {\rm cm}^2.
\end{eqnarray} 
Note that the string length scale $l_s$ determines the 
natural size of the $D$-brane as seen by an incoming string.
For each scattering event there will be a few
highly-energetic photons emitted due to the mechanism
described above.
This implies that, in order of magnitude,
the total number of emitted photons is about the same
as that of incident particles. Assuming that $\ell \sigma n_D $ is small,
it follows that the flux of the emitted photons will be roughly of order
$J_{he}\ell\sigma n_D$. Restricting ourselves from now on to the value
$g_s=10^{-2}$, we obtain  $\sigma \sim 10^{-54}
{\rm cm^{2}}$.

The emerging photons of the first particle/$D$-brane scattering
carry energies less than the threshold energy,
hence, when such photons strike another $D$-particle will be captured.
If $\ell \sigma n_D $ is small,
the final photon flux will still be given by
\begin{eqnarray}\label{fluxgamma}
   J_\gamma = J_{\rm he}\ell \sigma n_D
\end{eqnarray} 
since in this case 
$e^{-\ell \sigma n_D} \simeq 1 $.

As already mentioned, a good candidate for the 
highly energetic particles with energies above 
the threshold (\ref{threshold}), 
are neutrinos emitted from the 
GRB's. 
According to studies of the last few years~\cite{neutrino}, neutrinos of such 
energies can be
emitted from the fireball due to the muon decay that follows
the pion production, due to the interaction of fireball protons with
afterglow photons. The energy of neutrinos of appreciable
flux is about two orders of magnitude less than that of 
protons, which is supposed to be the highest energy of the  
ultra-high-energy cosmic rays (UHECR). Then,  
we expect to have neutrinos of flux of order, 
roughly, $10^{-9} {\rm GeV}/{\rm cm^{2}sec}$ at energies $10^{18}$ 
to $10^{19}$eV, 
depending on the energy and neutrino species~\cite{neutrino}.

If one assumes a uniform distribution of the $D$-particle defects
inside large regions in space, then a simple calculation shows that,
in order to explain the observed events, one needs a
phenomenologically unacceptable total mass of $D$-particles.
It is, therefore, necessary to 
assume that the latter form collections of very high 
density, whose concentration in space, however, is low.
We next assume that such collections can produce radiation flux of order
of that of the high energy cosmic rays,
$J_{\rm CR} \sim 10^{-20}/{\rm cm^{2}~sec~sr}$, and
that the observed UHECR flux is of the same order
as the photon flux (\ref{fluxgamma}). If there are $N$ such ensembles 
of $D$-particles
in a volume around 
Earth, of average radius in the order of 10 Mpc, one has
\begin{eqnarray} \label{relformula}
&~& J_{\rm CR} \sim  NJ_\gamma \sim NJ_{\rm he} \ell \sigma n_D =
Nc n_{\rm he} \ell \sigma n_D\sim \nonumber \\ 
&~& \sim N c n_{he}\sigma N_D/\ell^{2}
\end{eqnarray} 
where $N_D$ denotes the average number of branes in one of 
these $D$-brane-collections,
$c$ is the speed of light in vacuo,
and $n_{\rm he}$ is the concentration
of the high-energy particles, assumed to be neutrinos from
GRB's . Their flux, given above, implies a number density of high energy
particles $n_{he} \sim 10^{-28}/{\rm cm}^{3}$.
 Then,
\begin{eqnarray} \label{nepin}N N_D \sim \ell^{2}10^{48}/{\rm cm}^{2}. \end{eqnarray}  
Using the fact that $M_D=10^{-9}{\rm gr}$ for the given string scale and 
coupling, one obtains 
the total $D$-particle mass contained in the 10Mpc-volume as:
\begin{eqnarray} \label{totalmass}M_{total} \sim 10^{6} \ell^{2}/{\rm cm}^{2}~M_\odot \end{eqnarray} 
These
formulae show that in the limit of very small $\ell$ the total
number and mass of branes needed can be very small. We have then 
to determine the lower limit of the linear dimension of the brane
collections $\ell$. First we must take into account 
a few other requirements.
As we have stressed earlier, it is important that the scattering rate of
high energy particles passing through the brane ensembles
is small, so that 
\begin{equation} 
\ell \sigma n_D\sim \sigma N_D/ \ell^{2} \ll 1.
\label{ineq}
\end{equation} 
Also, in order to be able to treat the $D$-particle collections as distinct
spots of scattering, whose internal density is unrelated to their
concentration in space, we must have $n_D \gg \frac{N}{10 ~{\rm Mpc-volume}}$. 
Finally ,
the Schwarzschild radius has to be much smaller than $\ell$, so
that one is far away from the black hole limit. Then, it turns out that 
the strongest condition 
is (\ref{ineq}), and the last two are 
trivially satisfied.
Since there is no independent lower limit for
$N_D$, we may take for concreteness $N_D \sim 10$ assuming
that such a configuration forms a stable bound state of $D$-particles, 
an issue that we will discuss later on.
Setting $\ell$ to be ten orders of magnitude larger than the saturated 
limit, we then get
$\ell\sim 10^{-16}$ cm, which gives a value for the total number of
brane ensembles in the 10 Mpc-volume $N\sim 10^{16}$. These imply
a total $D$-brane mass in that volume of order 
$10^{-26} M_\odot \sim 10^{4} ~{\rm kgr}$. These numbers are quite sensitive in
changes of the parameters, but there are large regions of the
parameter space where the results are reasonable . The total mass
can be very small and hence it does not contribute
to the dark matter, although it would be a very good candidate for it, 
given that branes are ``bright'' only when very high energetic particles
are passing nearby.

The conclusion of the above, rather indicative, results is 
that a relatively
small number of $D$-particles is needed to produce an observable effect, if
our assumptions are valid. Then, an observed excess of photons in a certain 
energy channel of the high energy cosmic rays 
would imply the existence of $D$-particles, thereby 
fixing the value of the product $g_sM_s$ of the fundamental
string parameters. The above estimates provide
information only about a portion of the total population of $D$-particles
in the universe, assuming the latter exist, which contributes 
to an observable radiation.

It should be remarked that the above considerations 
were based on the assumption that the maximum energy 
of the incident particles (neutrinos or other species) 
is of order $10^{19}$ eV. Relaxing this, by allowing 
the maximum energies to be higher, e.g. of order 
$10^{21}$ eV~\cite{neutrino},
would imply that the above scenario could also provide 
a mechanism for production of UHECR beyond the 
GZK cut-off~\cite{gzkcutoff}. For this to happen, of course, 
the location of the brane defects must lie within the respective
mean free paths. However, we must note that 
these considerations 
cannot offer an explanation of the complete set of the observed UHECR 
events~\cite{gzkobs}. The so-produced UHECR events, if any,  
constitute only a contribution to this set, 
given that our mechanism produces only 
photons with energies in a rather narrow interval.

\section{On the Stability of Clusters of $D$-Particles in the Universe}

For the validity of the above scenaria it is crucial 
that the populations of the $D$-particle defects 
are {\it stable}. Naively, since the $D$-particles 
are very massive,
one expects that at large concentrations, such as the ones above,
they will collapse to form
black holes or at least to produce very strong gravitational fields,
which would jeopardize our scenario. 
The problem of preventing such a collapse
is equivalent to that of stabilizing $D$-branes, which
at present is an important open issue.
One expects that, in general, 
such stability mechanisms would 
imply a certain distance scale among 
the constituent $D$-particles in a collection, 
as well as restrictions
on the number of branes in the ensemble and probably 
on the total number of branes in the universe, something 
which most likely cannot be explained by the dynamics of their
interactions alone.

However, there are recent attempts in the string literature
in which -under certain circumstances- one has a {\it no force
condition} among the $D$-particles. Usually this is the
case of sufficient spacetime supersymmetries, where the
$D$-particle states are viewed as specific states, saturating the
so-called
Bogomol'nyi-Prasad-Sommerfield (BPS) bound~\cite{dual}.
However, it has been pointed out~\cite{dual2}
that one may obtain {\it stable} non BPS $D$-brane states
by considering appropriate combinations of BPS branes,
in such a way that spacetime supersymmetry is not preserved.
From our point of view we are interested in such stable non-BPS
non-supersymmetric $D$-particles.

Such states
can be viewed as solutions of certain string theories, which are
connected with the phenomenologically relevant string theories,
like Heterotic String, assumed to be living on
the four-dimensional spacetime
(after appropriate compactification),
by virtue of certain {\it
duality symmetries}~\cite{dual,dual2}.
One such theory is type IIB string theory.
In general, $D$-particle/$D$-particle scattering in type IIB string theory
has been studied in the literature~\cite{typeiib}. Although the
construction of \cite{typeiib}
refers to a specific string model involving a particular orbifold
compactification~\cite{dual3}, which may not be necessarily
realized in our situation, however we find it generic enough
so as to consider it as a prototype for the $D$-particle
stability mechanism
we need here.

The main idea behind the stability mechanism via the no force condition
in this model for $D$-particles
lies on the fact that there are additional vector interactions,
$U_D(1)$,
associated with the specific way of
constructing the non-BPS state.
We shall not explain the details here, but we refer the interested
reader to the relevant literature~\cite{dual2,typeiib,dual3}.
Such $U_D(1)$ interactions should not be confused with observable interactions
of ordinary matter in our construction. The latter may be assumed {\it neutral}
under the $U_D(1)$, which thus characterizes only the $D$-particles
in the ensemble, which are then 
assumed `electrically' charged under this $U_D(1)$, and in principle
may have both positive and negative charges. The positive
charges characterize, say, the $D$-particles, whilst the negative charges
the anti-D particles,
denoted by ${\overline D}$.
The $U_D(1)$  interaction is therefore
{\it repulsive} among $D$-particles (or ${\overline D}$-particle), 
and {\it attractive} among $D$-particle-${\overline D}$-particle.

Once we admit both kinds of charges
there  arises the issue as
to how ``polarized'' $D$-particle collections, characterized by
a significant 
excess of (positive or negative)  $U_D(1)$ charge, have been formed
in the Galaxies today. This
has not been resolved as yet, neither
will be
the topic of the present work. For our purposes here
we merely conjecture that, somehow, after the
Big Bang, such polarized regions emerged inside ordinary matter,
in such a way that the overall net $U_D(1)$ charge of the
Universe
is zero. An alternative scenario would be
that ${\overline D}$-particles behaved in similar
way as antiparticles in ordinary matter, and hence the Universe
today consists in its overwhelming majority of $D$-particles
only. This, would then trivially solve
the above-mentioned problem of ``polarization''.

In general, the no force condition may be a property valid at
all distances. This is the case of supersymmetric $D$-particles that saturate
the BPS bound. However,
in the orbifold construction of \cite{typeiib}, the no force condition
among the non BPS $D$-particles
is found to occur at large scales $r$, as
compared with the string length $l_s$, where notably (target space) effective
low-energy string action methods are 
applicable. This feature 
is exclusive of a critical-radius orbifold compactification.

An important point, however, which should be stressed here 
is that the construction of
\cite{typeiib} ignored higher-string loop effects.
The incorporation of such effects results in general
in the destruction of the no force condition~\cite{loops},
although under some circumstances it may be made valid up
to one loop. The effects
of string-loop resummation cannot be answered at present, and
hence the issue of the no force condition at a {\it non-perturbative}
string level is still open.

\section{Embedded Recoiling $D$-Particles in the Universe, \\
Cosmological Evolution and Particle Horizons}

In the previous sections we have demonstrated the dynamical
creation of event horizons around a recoiling $D$-particle 
defect, as a result of the induced distortion 
of the surrounding spacetime, due to back reaction effects. 
The original spacetime, in which the defect was embedded
had been assumed flat and static (at global scales). 
The perturbative analysis 
presented above is valid only for times large
after the impulse, and for distances sufficiently away 
from the defect, i.e. outside its Schwarzschild radius, 
so that induced space-time curvature effects 
due to the massive nature of the $D$-particle can be safely ignored.

An interesting question is how the presence of recoiling $D$-particles
in a cosmological Universe, of Friedman-Robertson-Walker 
type, affects the cosmological evolution.
This question acquires important physical significance, in view 
of our previous `phenomenological' analysis. 
This issue has been examined in detail in \cite{gravanis}, and we shall
not repeat the details here. 
For completeness, in what follows we shall review 
briefly the most important arguments and results.

Let us consider a $D$-particle, located (for convenience) at the origin 
of the spatial coordinates of a {\it curved} 
four-dimensional space time, which at 
a time $t_0$ experiences an impulse. 
In a $\sigma$-model framework, the trajectory of the $D$-particle 
$y^i(t)$, $i$ a spatial index,   
is described by inserting the following
vertex operator
\begin{equation} \label{path2}
V = \int _{\partial \Sigma} G_{ij}y^j(t)\partial_n X^i 
\end{equation}
where $G_{ij}$ denotes the spatial components of the metric, 
$\partial \Sigma$ denotes the world-sheet boundary, 
$\partial _n$ is a normal world-sheet derivative, 
$X^i$ are $\sigma$-model fields obeying Dirichlet boundary conditions 
on the world sheet, 
and $t$ is a $\sigma$-model field obeying Neumann boundary conditions
on the world sheet, whose zero mode is the target time.

This is the basic vertex deformation assumed 
to describe the motion of a $D$-particle in a curved geometry
to leading order at least, where spacetime back reaction 
and curvature effects are assumed weak.
Such vertex deformations may be viewed as a 
generalization of the flat-target-space case (\ref{recoilop}).
For times long after the event,   
where our perturbative approach is valid, 
the trajectory $y^i(t)$ will be that of free motion
in the 
curved space time under consideration. In the flat space time 
case, this trajectory was a straight line~\cite{kogan,szabo}, 
and in the more general 
case here it will be simply the associated {\it geodesic}.

In \cite{gravanis} we have dealt with space times 
of Friedman-Robertson-Walker (FRW) 
form:
\begin{equation}\label{rwmetric}
ds^2 = -dt^2 + a(t)^2 \sum_{i=1}^{3}(dX^i)^2  
\end{equation}
where $a(t)$ is the FRW scale factor. We shall work with 
expanding FRW space times with scale factors 
\begin{equation}\label{frwscale}  
a(t) =a_0 t^p, \qquad p \in R^+ 
\end{equation} 
where the times $t \gg t_0$, i.e. much later 
after the moment of impulse.

With initial conditions $y^i(t_0)=0$, and $dy^i/dt (t_0) \equiv v^i$, 
one easily finds that, for long times $t \gg t_0$ after the event,  
the path $y_i(t)$, which is a 
solution of the appropriate geodesic equations
for the spacetime (\ref{rwmetric}),(\ref{frwscale}), acquires the form: 
\begin{equation}\label{pathexpre} 
y^i(t) =\frac{v^i}{1-2p}\left(t^{1-2p}t_0^{2p} - t_0 \right) + 
{\cal O}(t^{1-4p}), \qquad t  \gg t_0 
\end{equation} 
In that case the deformation (\ref{path2}) becomes~\cite{gravanis}:
\begin{equation}\label{path} 
V=\int _{\partial \Sigma} a_0^2 \frac{v^i}{1-2p}\Theta_\epsilon (t-t_0)\left(tt_0^{2p}-t_0t^{2p}\right)
\partial_n X^i 
\end{equation} 
where $\Theta_\epsilon (t-t_0)$ is the regulated 
Heaviside step function (\ref{regultheta}),
expressing an instantaneous action ({\it impulse}) 
on the $D$-particle
at $t=t_0$.

The case $p > 1$ corresponds to an {\it accelerated }
Universe, ${\ddot a(t)} = p(p-1)a(t) >0$,
which suffers from the problem of a cosmological horizon, 
in the sense that the quantity
\begin{equation}\label{cosmohorizon} 
\delta \propto \int _{t_0}^\infty \frac{dt}{a(t)} =  
\int _{t_0}^\infty \frac{dt}{a_0t^p} < \infty, \qquad {\rm for}\, p>1
\end{equation}
is {\it finite}.  
This presents a problem in defining asymptotic states, and hence 
a proper $S$-matrix~\cite{accel}. Hence, as mentioned previously,
this constitutes 
a challenge for critical string theory~\cite{accel}. 

However, as became clear from our analysis above, non-critical 
string theory, with the target-time being 
identified as the Liouville field~\cite{emn},
is not an $S$-matrix theory~\cite{emnsmatrix}.
In such a case there is a well-defined superscattering 
matrix \$ which connects asymptotic density matrices, 
rather than pure states. 

However, there may be a possibility, which we shall point out 
in this section, which allows for a relaxation of a non-critical
string model to a critical one, thereby allowing for 
a proper asymptotic (far future) definition of pure states.
As we have mentioned previously, this is dictated 
by the fact that the anomalous dimension of the non-conformal 
recoil deformations (\ref{anom}) is running
with the time/Liouville-field and becomes zero asymptotically in time,
thereby implying an approach to a fixed (equilibrium) conformal point.

In the specific cosmological model at hand, 
this has been demonstrated 
in \cite{gravanis}, and will be reviewed below.
Specifically, we shall show that 
the recoil of the $D$-particle induces, via the associated non-criticality
of the string theory involved, a sufficient distortion in the space 
time global geometry so as to remove the cosmological 
Horizon and thus stop the cosmic acceleration. 
This phenomenon presumably will be enhanced by the presence 
of populations of $D$-particle defects, like the ones 
assumed in previous sections. 

To this end, we first recall~\cite{gravanis} that, from a world-sheet 
view point, the operators (\ref{path}),(\ref{rwmetric}) 
obey a higher order (determined by the exponent $p$) 
logarithmic algebra~\cite{lcft}, and are relevant 
operators in a renormalization-group sense, 
with anomalous dimension of the form (\ref{anomd}).
They are, therefore, in need of Liouville dressing. 
This is similar in spirit 
to their flat-space counterparts (\ref{recoilop}),
with the important physical difference that here the 
logarithmic operator pairs do not describe 
Galilean positions and velocities but rather
cosmic velocities and acceleration. 
The logarithmic nature of the deformations imply that they can 
still be classified by conformal blocks, and hence 
are `good' deformations from a conformal-field theory viewpoint.
This makes them acceptable backgrounds of string theory.
However, as a result of their relevant renormalization-group 
nature, the
resulting $\sigma$-model is non-critical. This is in agreement
with the fact that such deformations describe non-equilibrium
processes in the Universe. 

Dressing the operators with the Liouville mode, to 
restore conformal invariance, we have
\begin{eqnarray} \label{bulkop} 
V_{L, {\rm bulk}} = \int _{\Sigma} e^{\alpha_i \phi} 
\partial_\alpha \left(y_i(t)\partial^\alpha X^i \right), \qquad 
\alpha_i = -\frac{Q}{2} + \sqrt{\frac{Q^2}{4} + (2-h_i)}
\end{eqnarray} 
where $h_i$ is the conformal dimension of the bulk operator.
Above we have followed the method of dressing the 
non-critical operator, having it first expressed as a 
world-sheet bulk total derivative operator. 
One can show~\cite{gravanis} that completely
equivalent results are obtained if one dresses directly
the boundary operator (\ref{path}).

For this type of non-critical recoil models
the central charge deficit $Q$ 
has the form~\cite{gravanis}: 
\begin{equation} 
Q^2 = Q_0^2 + {\cal O}(\epsilon^2)
\end{equation}
There are two cases, depending 
on the value of $Q_0$:(i) $Q_0 \ne 0$, and (ii) $Q_0 =0$,
in which case the recoil is the only source of non-criticality 
in the model.  The case (i) is a feature of cosmological models
where there may be, in general, additional sources of non-criticality, as 
compared with the flat-space time case discussed in previous sections,
which is characterized by $Q_0=0$. 
This implies that the gravitational anomalous dimensions are 
of order: 
$\alpha_i \sim \epsilon^2$ if $Q_0 \ne 0$, and  
$\alpha_i \sim \epsilon$ if $Q_0 = 0$.

Consider first the case $Q_0 \ne 0$. 
In this case $\alpha_i \sim \epsilon^2$, 
$\phi_0 \sim \epsilon^{-2}$ and hence 
$\alpha_i \phi_0 \sim \epsilon t = {\rm const}$.
We identify now the Liouville direction $\phi$ 
with that of the target 
time~\cite{emn,recoil}. Given that $t \sim 
\tfrac{1}{\epsilon}$ this implies that 
$\phi \sim t^2$.    
Under this identification we observe~\cite{gravanis}
that 
the terms $e^{\alpha\phi_0}$, and  
the exponential 
factors $e^{-\epsilon~(t-t_0)}$ appearing in the regulated
$\Theta$ functions~\cite{kogan}
are all of order one. 

From these considerations one obtains an induced non-diagonal metric element 
$G_{0i}$: 
\begin{eqnarray} 
G_{0i}d\phi dX^i \simeq -v_i~\alpha_i t^{2p}d(t^2)dX^i~, 
\qquad t \gg t_0
\end{eqnarray} 
Under the fact that one 
identifies $\epsilon^{-1} =t-t_0 \sim t \gg t_0$, 
the non-diagonal element of the spacetime metric becomes:
\begin{eqnarray}\label{ndmetric2}
G_{0i} \sim v_i t^{2p-1} 
\end{eqnarray} 
We remind the reader that we analyze here the case with horizon, 
which implies $p > 1$.  

It is convenient now to diagonalize the metric, which implies
the following line element
\begin{eqnarray} \label{nontrans}
ds^2 = -\frac{v_i^2}{a_0^2}t^{2p-2}dt^2 + 
a_0^2t^{2p}(dX^i)^2 
\end{eqnarray} 
By redefining the time coordinate to $t'=\frac{\sqrt{{v}^2_i}}{a_0 p}t^p$ 
one obtains the induced line element:
\begin{eqnarray}\label{rwfinal} 
ds^2 = -(dt')^2 + \frac{a_0^4~p^2}{{v}_j^2}(t')^2~(dX^i)^2, \qquad t \gg t_0  
\end{eqnarray} 
From (\ref{cosmohorizon}), we thus observe that the induced metric  
has {\it no horizon}, and no cosmic acceleration. 
In other words a recoiling $D$-particle, 
embedded in a space time which initially appeared to have an horizon,
back reacted in such a way so as to remove it! Equivalently,
we may say that  
recoiling $D$-particles are consistent only in spacetimes 
without cosmological horizons. 

Similar conclusions are reached in the case $Q_0 =0$. 
In that case, $\phi/Q \sim \epsilon^{-2}$, 
for reasons associated with 
the normalization of the Liouville $\sigma$-model kinetic term, 
as explained 
previously (\ref{areaphi}),  
and since $Q \sim \epsilon \sim \tfrac{1}{t}$ 
in that case (cf. (\ref{centralcharge})), 
one 
has that $\phi \sim t$. Again, the exponential terms 
$e^{\alpha_i \phi}$, $\alpha_i \sim \epsilon$, 
and those coming from the 
regulated $\Theta_\epsilon (t)$ 
are of order one. 
Evidently, the induced non-diagonal
metric has the same form (\ref{ndmetric2}) 
as in the case with $Q_0 \ne 0$,
and one can thus repeat the previous analysis, implying removal 
of the cosmological horizon and stopping of cosmic acceleration.

The reader must have noticed that the same conclusion is 
reached already at the level of the metric (\ref{nontrans}),
before the time transformation, once one interprets the coefficient 
of the $(dt)^2$ as a time-dependent light velocity.
The fact that such situations arise `suddenly', after a time moment $t_0$, 
might prompt the reader to draw some analogy 
with scenaria of time-dependent 
light velocity, involving some sort of phase transitions 
at a certain moment in the (past) history of our Universe~\cite{moffat}.
In our case, as we have seen,  
one can perform (at late times) 
a change in the time coordinate in order to arrive at a RW metric 
(\ref{rwfinal})~\footnote{At this point  
we would like to 
draw the reader's attention to the fact that such transformations 
depend on the recoil velocity, and thus on the energy content
of the matter incident on the D-particle. 
On the other hand, it is not clear how one could extend 
the results of the present work to 
space-time  
`foam' situations~\cite{recoil}, in which  
several incident 
particles interact with collections of $D$-particles,
which are virtual quantum excitations of the string/brane vacuum.
Hence it might be 
that in such cases  
one cannot perform simultaneous transformations to diagonalize the metric,
thereby obtaining non-trivial refractive indices
of the type discussed in~\cite{sarkar,recoil}, implying time delays
in the arrival time of massless particles 
which increase 
with increasing energy. This should be contrasted to the 
thermal refractive indices discussed in section 7 above. 
Such issues fall beyond the scope of the present article.}. 

{}From a field-theoretic point of view, 
the removal of the horizon would seem to imply 
that one can define asymptotic states and thus a proper $S$-matrix.  
However, in the context of Liouville strings, with the Liouville mode 
identified
with the time~\cite{emn}, 
there is no proper $S$-matrix, independently of the existence 
of horizons~\cite{emnsmatrix}. 
As we have already discussed, this has to do with the structure
of the correlation functions of vertex operators in this construction,
which are defined over steepest-descent 
closed time-like paths in a path-integral formalism,
resembling closed-time paths of non-equilibrium 
field theories~\cite{emn,emnsmatrix}. In such constructions one can define
properly only a (non factorizable) 
superscattering matrix \$$\,\ne S\,S^\dagger$.  
Neverthelss, the removal of the cosmic horizon asymptotically, 
and the stopping of the acceleration of the Universe, 
are compatible with the approach of 
our non-critical string model to a (world-sheet) renormalization-group
fixed point. The latter is an {\it equilibrium} point, 
at which the theory becomes a critical string. 
It is for this reason that one can define proper 
asymptotic (far future) states.

As a final remark, we would like to point out that
in models where the induced non-criticality 
is an exclusive feature of the impulse destortion (\ref{path}), 
the induced four-dimensional 
vacuum energies are given by the central charge deficit,
which asymptotically in time behaves as 
(c.f. (\ref{centralcharge}),(\ref{time})):
\begin{equation}\label{quint} 
\Lambda_{4d} = Q^2 \sim \frac{1}{t^2}
\end{equation}
Such a relaxing-to-zero vacuum energy is compatible with 
recent observations, and in the above model is 
due exclusively to gravitational recoil degrees of freedom.
The dilaton field has been assumed constant.  

On the other hand, 
in certain non-critical cosmological models 
with central charge deficits $Q^2 = Q_0^2 + \dots $,
with $Q_0={\rm constant}  \ne 0$, 
and the $\dots $ denoting terms that go to zero as $t \to \infty$, 
there may be 
non-trivial
time-dependent dilaton fields, $\Phi (t) \sim -{\rm log}(t)$,
for $t \to \infty$.
Models with such a behaviour are discussed explicitly
in \cite{emnsmatrix,georgal}, where we refer the reader 
for details. Here we simply mention that  
such dilaton fields may operate as quintessence fields,
in the sense of yielding 
four-dimensional relaxing-to-zero vacuum energies 
of the form: 
\begin{equation}
\Lambda_{4d} = e^{2\Phi(t)}Q_0^2 \, \to \, \frac{Q_0^2}{t^2} 
\end{equation} 
where the dilaton exponential factors appear as a result 
of the fact that we work in the Einstein frame
of string effective actions, in which the Einstein curvature
term has the canonical normalization~\cite{GSW}. 

We therefore see that Liouville strings are compatible
with current cosmological observations. 
Of course we do not claim that there are no
conventional explanations of the currently observed
acceleration of the Universe, but we think that the above ideas
are interesting enough to encourage further studies of 
non-critical string cosmological models.  

\section{Instead of Conclusions}

In this work we have studied the formation of bubbles
as a result of scattering of closed strings off $D$-particles
embedded in a four dimensional spacetime. 
Such configurations may be thought of as a trivial case of 
intersecting branes, provided one adopts the modern view point
that our four-dimensional world is a $D3$-brane. 
In this sense, the string scale $M_s$, which enters our 
calculations, is not necessarily the same with the 
four-dimensional Planck scale, and as we have seen this played
an important r\^ole in our analysis, as it allowed us to 
work at distances sufficiently far from the Schwarzschild radius
of the $D$-particle, for finite $g_s$ string couplings.

An interesting feature of our approach 
is the entropy production, which we associated
with information carried away by the recoil degrees of freedom. 
This latter feature may have cosmological implications for 
mechanisms of entropy production in the early universe, where 
we expect the density of $D$-particles to be significant, and hence 
the probability of scattering with closed strings important.
The fact that highly energetic charged particles
can escape the bubble, with the simultaneous release of radiation,
which can also escape and thus is in principle 
observable, is interesting, and might 
imply important phenomenological constraints on the 
order of the density of the $D$-particles in the Universe today,
in the way discussed in section 8.
An important feature for all such scenaria, which is yet-to-be established
is the stability of the relevant population of $D$-particles. 
This issue is an open issue at present, 
and is exclusively a non-perturbative feature of 
string theory, which is beyond our control at present.

A comment we wish to make at this point concerns the 
impossibility of the extension of the above 
analysis, with the specific choice of fields,
to higher-dimensional 
spacetimes, of spacetime dimension $d>4$.
This comes about because, under the 
simplest form for the tachyonic mode $T = {\rm ln}r$,
consistent with   
Einstein's equations 
to order ${\cal O}(\alpha')$,  
one obtains 
the following form for the tachyonic-mode potential
$V(T) \sim 1/r^2$.  
On the other hand, the 
equation of motion for the mode itself, consistent 
with the above form for $T$,  
demands that $dV(T)/dT =-(d-2)V(T)$. 
Clearly, this is consistent  
only {\it in d=4} spacetime dimensions. 
Hence, despite the fact that in our approach we have 
restored Lorentz invariance, 
we still obtain a special r\^ole of $d=4$, 
as in the 
Lorentz violating scenario of \cite{recoil},
where the sacrifice of Lorentz invariance in the 
sense of an explicitly 
space-dependent vacuum energy,
lead also to a selection of $d=4$.
If this feature survives the inclusion of the complete string 
matter multiplets, something which is not clear to us at present, 
then, it might imply that a {\it recoiling} 
Liouville $D$-particle cannot be embedded in (intersect with)
a target spacetime
(viewed itself as a $D$-brane), consistently 
with the $\sigma$-model conformal invariance, 
unless 
its dimensionality is $d=4$. At present we consider the issue 
only as a mathematical curiosity of the specific
effective field theory at hand, and 
we do not attribute to it further physical 
significance. However, surprises cannot be excluded.

Another important result, described above, 
concerns a potential physical r\^ole
of embedded $D$-brane defects on the Cosmological 
Evolution of a Friedman-Robertson-Walker Universe.
It has been argued that the presence of 
a recoiling $D$-particle  
results in the 
removal of cosmological horizons and the eventual stopping of the 
Cosmic acceleration. 
From a field-theoretic
view point this would allow 
a proper definition of asymptotic states in such models. 
Moreover one obtains a 
relaxing to zero vacuum energy, compatible with current 
astrophysical observations.  

The above considerations pertain strictly to the case where the 
$D$-particle defects are considered as real. In quantum space-time 
foam situations such defects are emerging as {\it virtual excitations}
of the vacuum. In that case it is not clear how the present
results can be extended, given that such situations 
lie far beyond the perturbative regime assumed above. 
Nevertheless one might hope that, at least {\it qualitatively}, 
some of the stochastic properties discussed here, e.g. thermal
refractive indices and entropy growth, will survive the 
full quantum treatment, and will constitute characteristic features
of non-critical stringy models of space time foam.

What we have described above is an admittedly speculative, but nevertheless
interesting, at least in the authors' opinion, scenario on
possible physical aspects of a specific 
Liouville (non-critical) string model
of quantum gravity.
It is interesting that physical predictions from such models might lie within 
the sensitivity of immediate-future experimental facilities,
both terrestrial and extraterrestrial. 
The basic feature of non-critical string models is their 
instability, as a result of their non-equilibrium nature. 
As such,  these theories may be quite relevant for the 
physics of the Early Universe, and also may be responsible 
for (part of) observed extreme astrophysical phenomena 
of cosmological origin, 
whose evidence has recently started becoming overwhelming. 
Gamma Ray Bursters, Ultra-High-Energy Cosmic Rays and others, 
are among the most obvious candidates to search for new physics,
either within the (conventional) local field theories, or
in the (yet to be established) framework of modern approaches
to quantum gravity, like critical strings or even non-critical ones.

Admittedly, the ideas presented in this article 
are very speculative, 
and indeed may have nothing to do at the end with  
real Physics. However, the mathematical and logical consistency
of such unconventional models, and their effects, 
seems, at least to the authors, 
convincing enough so as not to discard them immediately.
Our basic point of view in this article, 
was that, if one adopts the modern 
approach of viewing our world as a domain wall
(brane), embedded in a higher dimensional 
(bulk) space time, then the possibility of having 
intersecting brane configurations cannot be logically excluded. 
Embedded $D$-particles in a four-dimensional domain wall is 
a special (and the simplest) case of intersecting branes.
As we have seen above, the fluctuations and scattering 
of the $D$-particles with other matter in the Universe 
may lead to physical effects, e.g. 
in connection with some extreme astrophysical
phenomena, which severely constrain
their populations.

It is actually very intriguing that stringy defects,
which at first sight are not expected to play any r\^ole 
in low-energy physics, may actually be partly responsible 
for some extreme astrophysical phenomena, whose observation 
became only recently possible, as a result of the 
enormous technological advances in 
both terrestrial and extraterrestrial instrumentation.  
It is always intellectually challenging, but also expected intuitively 
in some sense, to think of the vast Universe as 
being the `next-generation'
Laboratory, where ideas on the quantum structure of spacetime
may be finally subjected to experimental tests in the not-so-distant future.

\section*{Acknowledgements}

The work of E.G. is supported 
by a King's College London Research Studentship (KRS).
N.E.M. wishes to thank H. Hofer (ETH, Zurich and CERN) 
for his interest and partial support.

\end{document}